\documentclass[10pt]{article}
\usepackage{amssymb}
\usepackage{axodraw}
\usepackage{rotate}
\usepackage{epsfig}
\usepackage{graphicx}
\usepackage{wrapfig}
\usepackage{floatflt}
\begin{document}
\newcommand{\eqnzero}{\setcounter{equation}{0}} 
\newcommand{\mpar}[1]{{\marginpar{\hbadness10000%
                      \sloppy\hfuzz10pt\boldmath\bf#1}}%
                      \typeout{marginpar: #1}\ignorespaces}
\def\mnew{\mpar{\hfil NEW \hfil}\ignorespaces}
\newcommand{\bq}{\begin{equation}}
\newcommand{\eq}{\end{equation}}
\newcommand{\bqa}{\begin{eqnarray}}
\newcommand{\eqa}{\end{eqnarray}}
\newcommand{\baa}[1]{\begin{array}{#1}}
\newcommand{\eaa}{\end{array}}
\newcommand{\nll}{\nonumber\\}

\newcommand{\ip }[1]{u\left({#1}        \right)}    
\newcommand{\iap}[1]{{\bar{v}}\left({#1}\right)}    
\newcommand{\op }[1]{{\bar{u}}\left({#1}\right)}    
\newcommand{\oap}[1]{v\left({#1}\right)}            

\newcommand{\Litwo}{\mbox{${\rm{Li}}_{2}$}}
\newcommand{\alem}{\alpha_{em}}
\newcommand{\alsS}{\alpha^2_{_S}}
\newcommand{\ds }{\displaystyle}
\newcommand{\sss}[1]{\scriptscriptstyle{#1}}
\newcommand{\sla}[1]{/\!\!\!#1}
\def\el{e}
\def\mgn{mgn}
\def\mw {M_{\sss{W}}}
\def\mws{M_{\sss{W}^2}}
\def\mz {M_{\sss{Z}}}
\def\mh {M_{\sss{H}}}
\def\men{m_{\nu_e}}
\def\mel{m_e}
\def\mup{m_u}
\def\mdn{m_d}
\def\mmn{m_{\nu}}
\def\mmo{m_{\mu}}
\def\mch{mch}
\def\mst{mst}
\def\mtn{mtn}
\def\mta{mta}
\def\mtp{m_t}
\def\mbt{m_b}
\def\mf{m_f}
\def\mv{M_{\sss{V}}}
\def\srt{\sqrt{2}}
\def\qel{Q_f}
\newcommand{\sqrtL}[3]{\sqrt{\lambda\big(#1,#2,#3\big)}}
\newcommand{\vpa}[2]{\sigma_{#1}^{#2}}
\newcommand{\vma}[2]{\delta_{#1}^{#2}}
\newcommand{\af}{I^3_f}
\newcommand{\sqs}{\sqrt{s}}
\newcommand{\stw}{s_{\sss{W}}  }
\newcommand{\ctw}{c_{\sss{W}}  }
\newcommand{\stws}{s^2_{\sss{W}}}
\newcommand{\stwf}{s^4_{\sss{W}}}
\newcommand{\ctws}{c^2_{\sss{W}}}
\newcommand{\ctwf}{c^4_{\sss{W}}}
\newcommand{\bff}[4]{B_{#1}\big( #2;#3,#4\big)}             
\newcommand{\fbff}[4]{B^{F}_{#1}\big(#2;#3,#4\big)}        
\newcommand{\scff}[1]{C_{#1}}             
\newcommand{\sdff}[1]{D_{#1}}                 
\newcommand{\dffp}[6]{D_{0} \big( #1,#2,#3,#4,#5,#6;}       
\newcommand{\dffm}[4]{#1,#2,#3,#4 \big) }       
\newcommand{\tHmus}{\mu^2}
\newcommand{\epsh}{\hat\varepsilon}
\newcommand{\epsb}{\bar\varepsilon}

\newcommand{\chapt}[1]{Chapter~\ref{#1}}
\newcommand{\chaptsc}[2]{Chapter~\ref{#1} and \ref{#2}}
\newcommand{\eqn}[1]{Eq.~(\ref{#1})}
\newcommand{\eqns}[2]{Eqs.~(\ref{#1})--(\ref{#2})}
\newcommand{\eqnss}[1]{Eqs.~(\ref{#1})}
\newcommand{\eqnsc}[2]{Eqs.~(\ref{#1}) and (\ref{#2})}
\newcommand{\eqnst}[3]{Eqs.~(\ref{#1}), (\ref{#2}) and (\ref{#3})}
\newcommand{\eqnsf}[4]{Eqs.~(\ref{#1}), 
          (\ref{#2}), (\ref{#3}) and (\ref{#4})}
\newcommand{\eqnsv}[5]{Eqs.(\ref{#1}), 
          (\ref{#2}), (\ref{#3}), (\ref{#4}) and (\ref{#5})}
\newcommand{\tbn}[1]{Table~\ref{#1}}
\newcommand{\tabn}[1]{Tab.~\ref{#1}}
\newcommand{\tbns}[2]{Tabs.~\ref{#1}--\ref{#2}}
\newcommand{\tabns}[2]{Tabs.~\ref{#1}--\ref{#2}}
\newcommand{\tbnsc}[2]{Tabs.~\ref{#1} and \ref{#2}}
\newcommand{\fig}[1]{Fig.~\ref{#1}}
\newcommand{\figs}[2]{Figs.~\ref{#1}--\ref{#2}}
\newcommand{\figsc}[2]{Figs.~\ref{#1} and \ref{#2}}
\newcommand{\sect}[1]{Section~\ref{#1}}
\newcommand{\sects}[2]{Sections~\ref{#1} and \ref{#2}}
\newcommand{\subsect}[1]{Subsection~\ref{#1}}
\newcommand{\appendx}[1]{Appendix~\ref{#1}}
\def\Cmi{c_{-}}
\def\Cpl{c_{+}}
\def\spr{s^{'}}
\def\betap{\beta_{+}}
\def\betam{\beta_{-}}
\def\betapl{\beta^c_{+}}
\def\betami{\beta^c_{-}}
\def\klmi{k^{-}_1}
\def\klpl{k^{+}_1}
\def\betaf{\beta_f}
\def\ph{\phantom{-}}
\def\phph{\phantom{phantomphantom}}
\def\GF {G_{\sss F}}
\def\gw {\Gamma_{\sss W}}
\def\gz {\Gamma_{\sss Z}}
\def\stw{s_{\sss W}}
\def\ctw{c_{\sss W}}
\newcommand{\GeV}{\unskip\,\mathrm{GeV}}
\newcommand{\MeV}{\unskip\,\mathrm{MeV}}

\setcounter{page}{0}
\thispagestyle{empty}

\begin{flushright}
{\tt IFJPAN-V-2005-06\\
{\tt hep-ph/0506120} \\
      June 2005      
}
\end{flushright}
\vspace*{\fill}
\begin{center}

{\LARGE\bf SANCnews: Sector $ffbb$ \\[1mm]}
\vspace*{4mm}
{\bf D.~Bardin$^{a,b,*,\star}$, S.~Bondarenko$^{c}$, L.~Kalinovskaya$^{a,b,\star}$,
 G.~Nanava$^{a,d,*}$, \\[1mm]
L.~Rumyantsev$^{b}$, and W. von Schlippe$^{e}$}
\vspace*{4mm}

{\normalsize{\it 
$^{a}$ IFJ im. Henryka Niewodnicza{\'n}skiego, PAN           \\
       ul. Radzikowskiego 152, 31-342 Krak{\'o}w,            \\
                  on leave from \\
$^{b}$ Dzhelepov Laboratory for Nuclear Problems, JINR,      \\
        ul. Joliot-Curie 6, RU-141980 Dubna, Russia;         \\
$^{c}$ Bogoliubov Laboratory of  Theoretical Physics, JINR,  \\ 
        ul. Joliot-Curie 6, RU-141980 Dubna, Russia;         \\
$^{d}$ on leave from IHEP, TSU, Tbilisi, Georgia;            \\
$^{e}$ Petersburg Nuclear Physics Institute,                 \\ 
        Gatchina, RU-188300 St. Petersburg, Russia.}}
\vspace*{4mm}

\end{center}

\begin{abstract}
\noindent
In this paper we describe the implementation of processes $f_1 \bar{f}_1 ZZ \to 0$ and 
$f_1 \bar{f}_1 HZ \to 0$ into the framework of {\tt SANC} system.
The $f_1$ stands for a massless fermion $f$ whose mass is  kept non-zero only in arguments
of $\ln$ functions and $\to 0$ means that all 4-momenta flow inwards. The derived scalar
form factors can be used for any cross channel after an appropriate permutation of their 
arguments ($s,t,u$). We present the covariant and helicity amplitudes for these processes: 
for the former only in the annihilation channel $f_1\bar{f}_1\to ZZ$, while for the latter 
in annihilation $f_1\bar{f}_1\to HZ$ and decay $H\to Z f_1\bar{f}_1$ channels.
We briefly describe additional precomputation modules which were not covered
in the previous paper. For the processes $f_1\bar{f}_1\to HZ(ZZ)$ and decay
$H\to Z f_1 \bar{f}_1$ we present compact results of calculation of the accompanying 
bremsstrahlung and discuss exhaustive numerical results. 

As applications there are two types of the Monte Carlo generators for the process $H\to 4\mu$.
The first one is the generator based on a single resonance approximation for one of the $Z$
bosons.
The second one, exploiting the double resonance approximation, is not described in this article.
For the generator in the single approximation we present a short description.

Whenever possible, we compare our results with those existing in the literature.
For example, we present a comparison of the results for $H\to 4\mu$ decay with those obtained by
MC generator Prophecy4f.
\vspace{1mm}

{\tt SANC} client for version {\tt v.1.10} can be downloaded from servers at CERN 
{\it http://pcphsanc.cern.ch/ (137.138.39.23)} and Dubna
{\it http://sanc.jinr.ru/ (159.93.75.10)}.
\end{abstract}

\centerline{\it (Submitted to Computer Physics Communications)}

\vspace{4mm}

\footnoterule
\noindent
{\footnotesize \noindent
$^{\star}$ Supported in part by EU grant MTKD-CT-2004-510126,
  in the partnership with CERN PH/TH Division.
\\
Supported in part by INTAS grant $N^{o}$ 03-51-4007.
\\
$^{*}$ Corresponding author.\\
\phantom{$^{*}$}{\it E-mail addresses:} sanc@jinr.ru, bardindy@mail.cern.ch (D.~Bardin)}
\clearpage
\tableofcontents
\clearpage
\listoffigures
\listoftables    
\clearpage

\begin{center}
{\bf PROGRAM SUMMARY}
 \begin{itemize}

 \item{} {\it Title of program}: {\tt SANC}
 \item{} {\it Catalogue identifier}: ADXK\_v1\_1
 \item{} {\it Program summary URL:} http://cpc.cs.qub.ac.uk/summaries/ADXK\_v1\_1
 \item{} {\it Does the new version supersede the previous version?:} Yes
 \item{} {\it Reasons for the new version:} implementation of new processes; extension of 
an automatic generation of FORTRAN codes by the {\tt s2n.f} package onto many more processes; bug fixes
 \item{} {\it Summary of revisions:}
 \begin{itemize}
  \item{} implementation of light-by-light scattering and Compton scattering with one virtual proton
          into QED branch
  \item{} vast update of {\bf 2f2b} node in EW branch
  \item{} complete renovation of QCD branch
 \end{itemize}
 \item{} {\it Program obtainable from}: CPC Program Library, Queen's University of Belfast, N. Ireland
 \item{} {\it Designed for:} platforms on which Java and FORM3 are available
 \item{} {\it Tested on:} Intel-based PC's
 \item{} {\it Operating systems}: Linux, Windows
 \item{} {\it Programming languages used}: Java, FORM3, PERL, FORTRAN
 \item{} {\it Memory required to execute with typical data}: 10 Mb
 \item{} {\it No. of bytes in distributed program, including test data, etc.:} 
 \item{} {\it No. of bits in a word}: 32
 \item{} {\it No. of processors used}: 1 on {\tt SANC} server, 1 on {\tt SANC} client
 \item{} {\it Distribution format}: tar.gz
 \item{} {\it Nature of physical problem}: Automatic calculation of pseudo- and realistic
            observables for various processes and decays in the Standard Model of Electroweak
            interactions, QCD and QED at one-loop precision level. Form factors and 
            helicity amplitudes free of UV divergences are produced. For exclusion of
            IR singularities the soft photon emission is included.
 \item{} {\it Method of Solution}: Numerical computation of analytical formulae 
                                 of form factors and helicity amplitudes.  
                                 For simulation of two fermion 
                                 radiative decays of Standard Model bosons $(W^{\pm},Z)$
                                 and the Higgs boson a Monte Carlo technique is used. 
 \item{} {\it Restrictions on the complexity}: In the current version of {\tt SANC}
                                             there are 3 and 4 particle processes 
                                             and decays available at one-loop precision level.
\item{} {\it Typical Running time}: The running time depends on the selected process.
                                    For instance, the symbolic calculation of 
                                    form factors (with precomputed building blocks)
                                    for $H\to e^{+}e{-}Z$ process takes about 10 sec,  
                                    helicity amplitudes --- about 10 sec,
                                    and bremsstrahlung --- 1 min 10 sec. 
                                    The relevant s2n runs take about 2 min 40 sec, 1 sec 
                                    and 30 sec respectively.
                                    The numerical computation of decay rate for this process 
                                    (production of bencmark case 3 Table) 
                                    takes about 5 sec (CPU 3GHz IP4, RAM 512Mb, L2 1024 KB).
\end{itemize}
\end{center}

\clearpage

\section{Introduction}

In this paper we continue to describe the computer system {\tt SANC} {\em Support of 
Analytic and Numerical Calculations for experiments at Colliders}~\cite{Andonov:2004hi}
intended for semi-automatic calculations of realistic and pseudo-observables for various 
processes of elementary particle interactions at the one-loop precision level.
This is done in the spirit of the first description of the {\tt SANC} system
(see~\cite{Andonov:2004hi}  and references therein
which we recommend as a first acquaintance with the system).

Here we consider the implementation of several processes of $ffbb\to 0$ kind
(where $f$ stands for a {\em fermion}, $b$ for a {\em boson} of the Standard Model (SM),
while for concrete bosons we use $A$ for the photon and $Z,\;W^{\pm},\;H$)
One should emphasize also that the notation 
$ffbb\to 0$ means that all external 4-momenta flow inwards; this is
the standard {\tt SANC} convention
which allows to compute one-loop covariant amplitude (CA) and form factors (FF) only once 
and obtain it for a concrete channel by means of a crossing transformation.
The present level of the system is realized in version {\tt v.1.10}.
Compared to version {\tt v.1.00}, it is upgraded both physics-wise and computer-wise.
As far as physics is concerned it contains an upgraded treatment of 
$u\bar{d}\to l^{+}\nu_l$ and $d\bar{u}\to l^{-}\bar{\nu}_l$ processes 
(see Ref.~\cite{Arbuzov:2005dd}) and a complete implementation of
$F\to f+f_1+\bar{f^{'}_1}$ CC decays up to numbers and MC generators.
(Here $F$ and $f$ stand for massive fermions and $f_1$ and $\bar{f^{'}_1}$ for massless 
fermions of the first generation.)
Although the version {\tt 1.10} tree literally contains only $t\to b + l^{+} + \nu_l$ 
decay~\cite{Sadykov:2005xx}, any decay of the kind $F\to f+f_1+\bar{f^{'}_1}$ may be treated 
in a similar manner and we are going to implement them into the next versions. 
The complete description of these CC decays will be given elsewhere~\cite{Arbuzov:2007ke}.
Version {\tt 1.10} contains also the process $H\to f_1 \bar{f}_1 A$ 
in three cross channels~\cite{Bardin:2007wb} in EW branch,
$\gamma\gamma\to\gamma\gamma$ scattering~\cite{Bardin:2006sn}
and $l\,l\to\gamma\gamma^{*}$ in QED branch, as well as a new QCD branch~\cite{SANC-QCD}.

New in version {\tt 1.10} are also several $ffbb \to 0$ processes, to whose implementation
this paper is devoted. We describe here two of them: $f_1 \bar{f}_1 ZZ \to 0$ and 
$f_1 \bar{f}_1 HZ\to 0$, the latter one being used in two channels --- annihilation and decay.

In the annihilation channel, these processes were considered in the literature extensively
(see, for instance,~\cite{Denner:1988tv} and~\cite{Ciccolini:2003jy}--\cite{ffHZHL}), 
however, we are not aware of publications devoted to the $H\to Z f_1 \bar{f}_1$ decay.

These processes are relevant for $H$ search at LHC: the processes
$f_1 \bar{f}_1 \to ZZ$ are one of the  backgrounds while the one-loop
calculations of the decay  $H\to Z f_1 \bar{f}_1 $ was also used for an improved treatment 
of the decay $H\to 4\mu$ for an intermediate Higgs mass 
interval 130 GeV $\leq M_{\sss H} \leq $ 150 GeV, see section~\ref{generator}.

Furthermore, in the spirit of the adopted {\tt SANC} approach, all $2f2b \to 0$ processes can be
computed with off shell bosons thereby allowing their use also as building blocks for 
future studies of $5\to 0$ processes.

\begin{figure}[!h]
\[
\begin{array}{ccc}
  \vcenter{\hbox{
\begin{picture}(132,132)(0,0)
 \ArrowLine(55,115)(33,115)
 \ArrowLine(55,17)(33,17)
 \Photon(0,110)(55,110){2}{12}
 \Vertex(0,110){2.5}
 \ArrowLine(0,110)(-22,132)
 \ArrowLine(0,22)(0,110)
 \Vertex(0,22){2.5}
 \ArrowLine(-22,0)(0,22)
 \Photon(0,22)(55,22){2}{12}
 \ArrowLine(-16,0)(0,16)
 \ArrowLine(-16,132)(0,116) 
 \Text(-32,132)[lt]{\sf ii}
 \Text(42,118)[lb]{$p_4$}    
 \Text(42,5)[lb]{$p_3$}
 \Text(42,60)[lb]{$t$-channel}
 \Text(-15,106)[lb]{$\nu$}
 \Text(60,105)[lb]{vu=typeFU}
 \Text(60,18)[lb]{vd=typeFD}
 \Text(-42,137)[lb]{fu=typeIU}
 \Text(-42,-12)[lb]{fd=typeID}
 \Text(-15,17)[lb]{$\mu $}
 \Text(-4,123)[lb]{$p_1$}
 \Text(-4,3)[lb]{$p_2$}
 \Text(35,-13)[lb]{a)}
\end{picture}}}
& \qquad &
  \vcenter{\hbox{
\begin{picture}(132,132)(0,0)
 \ArrowLine(55,115)(33,115)
 \ArrowLine(55,17)(33,17)
 \Photon(0,110)(55,110){2}{12}
 \Vertex(0,110){2.5}
 \ArrowLine(0,110)(-22,132)
 \ArrowLine(0,22)(0,110)
 \Vertex(0,22){2.5}
 \ArrowLine(-22,0)(0,22)
 \Photon(0,22)(55,22){2}{12}
 \ArrowLine(-16,0)(0,16)
 \ArrowLine(-16,132)(0,116) 
 \Text(-32,132)[lt]{\sf ii}
 \Text(42,118)[lb]{$p_3$}    
 \Text(42,5)[lb]{$p_4$}
 \Text(42,60)[lb]{$u$-channel}
 \Text(-15,106)[lb]{$\nu$}
 \Text(60,105)[lb]{vd=typeFD}
 \Text(60,18)[lb]{vu=typeFU}
 \Text(-42,137)[lb]{fu=typeIU}
 \Text(-42,-12)[lb]{fd=typeID}
 \Text(-15,17)[lb]{$\mu $}
 \Text(-4,123)[lb]{$p_1$}
 \Text(-4,3)[lb]{$p_2$}
 \Text(35,-13)[lb]{b)}
\end{picture}}}
\end{array}
\]
\vspace*{-2mm}
\caption[Born $ffHZ(ZZ)\to 0$ diagrams, $t$ and $u$ channels]
        {Born $ffHZ(ZZ)\to 0$ diagrams, $t$ and $u$ channels.}
\label{Borntu}
\vspace*{-15mm}
\end{figure}
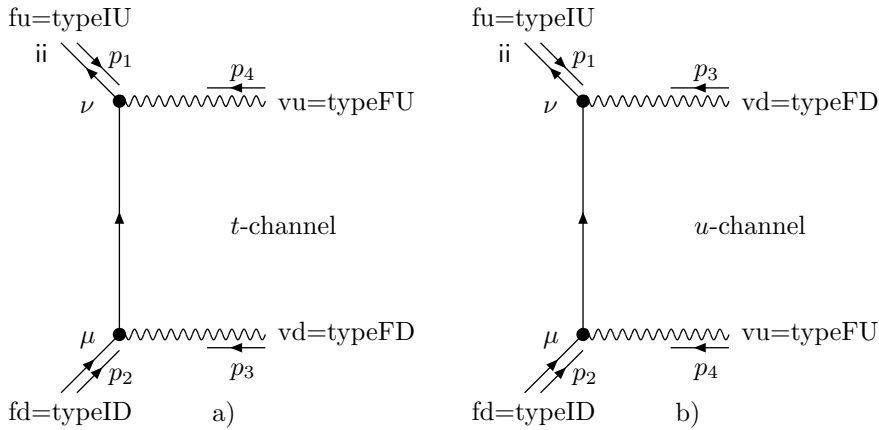

The process $f_1 \bar{f}_1 ZZ$ is very similar to the processes $ffb\gamma$ 
($b=\gamma,\,Z\,,H$) whose precomputation was described in detail in 
Ref.~\cite{Andonov:2004hi}. 
Its tree level amplitude is represented by two diagrams in $t$ and $u$ channels, 
Fig.~\ref{Borntu}.
\vspace*{-4mm}

\clearpage

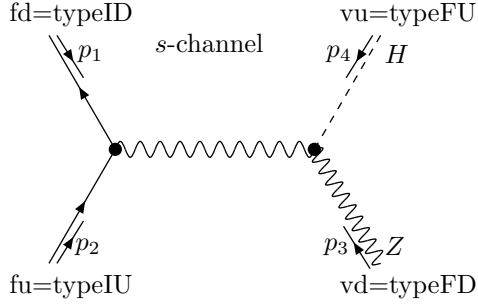
\begin{figure}[!t]
\[
\begin{picture}(125,86)(0,0)
  \Vertex(100,43){2.5}
  \DashLine(125,86)(100,43){3}
  \Photon(100,43)(125,0){3}{10}
  \Photon(25,43)(100,43){3}{10}
  \Vertex(25,43){2.5}
  \ArrowLine(0,0)(25,43)
  \ArrowLine(25,43)(0,86)
 \Text(-15,90)[lb]{fd=typeID}
 \Text(-15,-13)[lb]{fu=typeIU}
 \Text(110,90)[lb]{vu=typeFU}
 \Text(110,-13)[lb]{vd=typeFD}
 \Text(40,80)[lb]{$s$-channel}

 \ArrowLine(3,0)(13,16)
 \ArrowLine(3,86)(13,70)

 \Text(11,76)[lb]{$p_1$}
 \Text(10,3)[lb]{$p_2$}
 \Text(127,76)[lb]{$H$}
 \Text(127,3)[lb]{$Z$}
 \ArrowLine(122,86)(112,70)
 \ArrowLine(121,-2)(111,14)

 \Text(105,76)[lb]{$p_4$}
 \Text(104,3)[lb]{$p_3$}

\end{picture}
\]
\vspace*{-6mm}
\caption[Born $ffHZ\to 0$ diagrams, $s$ channel]
        {Born $ffHZ\to 0$ diagrams, $s$ channel.}
\label{Borns}
\vspace*{-2mm}
\end{figure}

Note, that in this and in the following figures, 
{fu=typeIU} etc. denote ``types'' of external particles, see Table 2 of~\cite{Andonov:2004hi}
as well as the discussion in the beginning of section~\ref{renormalization}.

\begin{floatingfigure}{65mm}
\hspace*{-5mm}
\includegraphics[width=65mm,height=115mm]{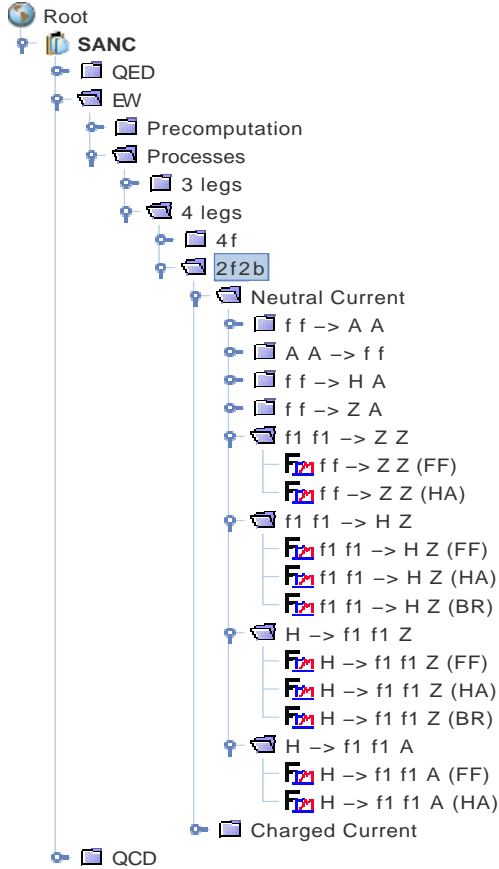}
\caption[New processes in the EW part]
        {New processes in the EW part.}\label{ProcEW11}
\end{floatingfigure}

For the process $f_1 \bar{f}_1 \to HZ$ these two diagrams do not contribute in the tree
approximation since fermion $f_1$ is considered to be massless. However, in this case
there exists an $s$ channel amplitude, Fig.~\ref{Borns},

All $ffbb$ processes are fully implemented at Level 1 of analytical calculations.
Several new modules which compute the contribution of the $bbb$ vertices to $ffbb\to 0$
processes are added to the precomputation tree, as well as three other modules relevant
for the $s$ channel diagram.

The modified ``Precomputation'' tree is shown in Fig.~\ref{PrecEW11} and discussed
in section~\ref{precomputation}. 

The modified branch {\bf 2f2b} for the ``Processes'' tree is shown in Fig.~\ref{ProcEW11}.
It contains four new sub-menus $f_1 \bar{f}_1\to ZZ$, $f_1 \bar{f}_1\to HZ$, 
$H\to f_1 \bar{f}_1 Z$ and $H\to f_1 \bar{f}_1 A$ which in turn are branched into scalar 
Form Factors (FF) and Helicity Amplitudes (HA) 
(two for the process $f_1 \bar{f}_1 HZ$ corresponding to two annihilation and decay channels) 
and the accompanying brem\-sstrahlung contributions (BR).

These processes are implemented also at Level 2, where the {\tt s2n.f} package produces the results
in the ``Semi Analytic'' mode 
(see Fig.~20 of Ref.~\cite{Andonov:2004hi}).
For the three decays we have relevant ``Monte Carlo'' generators which, however, are not yet 
implemented into the system. 

The paper is organized as follows:

In section~\ref{amplitudes} we describe the covariant (CA) and helicity amplitudes for three of 
four new $ffbb \to 0$ processes available in {\tt version 1.10}.

Section~\ref{precomputation} contains a brief description of new precomputation modules.
In section~\ref{renormalization} we describe in some more detail the renormalization procedure 
for the $f_1 \bar{f}_1 HZ\to 0$ process, {\em i.e.} calculation of FFs.
Section~\ref{brems} contains the results for the accompanying 
brem\-sstrahlung in the semi-analytic mode for two $f_1 \bar{f}_1 HZ\to 0$ channels.
Section~\ref{numerics} contains numerical results for the processes
$f_1 \bar{f}_1 \to HZ(ZZ)$ and decay $H \to Z f_1 \bar{f}_1 $.
Finally, section~\ref{generator} contains a brief description a
Monte Carlo generator for process $H \to 4 \mu$ in the single resonance approximation. 
The first results of numerical comparison with those of 
Prophecy4f~\cite{Bredenstein:2006rh}--\cite{Buttar:2006zd} are also presented.

\vspace*{-20mm}

\clearpage

\section{Amplitude Basis, Scalar Form Factors, Helicity Amplitudes\label{amplitudes}}
\subsection{Introduction}
In this section we continue the presentation of formulae for the amplitudes of
 $ffbb\to 0$ 
processes started in section 2 of Ref.~\cite{Andonov:2004hi}.
As usual, we begin with the calculation of CAs corresponding to a result 
of the straightforward computation of {\em all} diagrams contributing to a
 given process at 
the one-loop level. It is represented in a certain {\em basis of structures},
 made of strings 
of Dirac matrices and external momenta, contracted with polarization vectors
 of vector bosons. 
The amplitude is parameterized by a number of FFs, which we denote
by ${\cal F}$ with an index labeling the corresponding structure.
The number of FFs is by construction equal to the number of structures,
however for the cases presented below some of the FFs can be equal, so the number of
independent FFs may be less than the number of structures. 
For the existing tree level structures the corresponding FFs have the form
\bqa
{\cal F} = 1 + \frac{\alpha}{4\pi\stw^2} {\tilde{\cal F}}\,,
\label{born_like}
\eqa 
where ``1'' is due to the Born level and ${\tilde{\cal F}}$ is due to
 the one-loop level. 
As usual, we use various coupling constants:
\bqa
Q_f\,,\quad I^{(3)}_f\,,
\quad \sigma_f = v_f + a_f\,,\quad \delta_f = v_f - a_f\,,
\quad \stw=\frac{e}{g}\,,\quad 
\ctw=\frac{\mw}{\mz}\,,\quad \mbox{\it etc.}
\eqa

Given a CA, {\tt SANC} computes a set of HAs, denoted by 
${\cal H}_{\lambda_1 \lambda_2 \lambda_3\dots}$,
 where $\lambda_1 \lambda_2 \lambda_3\dots$ 
denote the signs of particle spin projections onto a quantization axis.

\subsection{$f\bar{f}\to ZZ$ process}
Here we present the CA of the process $f(p_2)\bar{f}(p_1)\to Z(p_3)Z(p_4)$
in the annihilation channel
\footnote{The other channels are unphysical in this case.}, see Fig.~\ref{Borntu}. 

It contains 10 left $(\gamma_{+})$ and 10 right $(\gamma_{-})$ structures:
\bqa
{\cal A}_{\sss ffZZ}
       &=& k_0 \,\Biggl\{
\Biggl[   \iap{p_1} \biggl(
       \sla{p_3} \gamma_{+} (p_1)_{\mu} (p_1)_{\nu} {\cal F}^{+}_1(s,t)
    +  \sla{p_3} \gamma_{+} (p_1)_{\mu} (p_2)_{\nu} {\cal F}^{+}_2(s,t)
\nll[-2mm] &&
\phantom{-\frac{i}{8\ctw^2} \,\Biggl\{ \Biggl[}
    +  \sla{p_3} \gamma_{+} (p_1)_{\nu} (p_2)_{\mu} {\cal F}^{+}_3(s,t)
    +  \sla{p_3} \gamma_{+} (p_2)_{\mu} (p_2)_{\nu} {\cal F}^{+}_4(s,t)
    +  \sla{p_3} \gamma_{+} \delta_{\mu\nu}         {\cal F}^{+}_5(s,t)
\nll[-2mm] &&
\phantom{-\frac{i}{8\ctw^2} \,\Biggl\{ \Biggl[}
    +  \gamma_{\mu}\sla{p_3}\gamma_{\nu} \gamma_{+} {\cal F}^{+}_6(s,t)
    +  \gamma_{\mu}\gamma_{+} (p_1)_{\nu} {\cal F}^{+}_7(s,t)
    +  \gamma_{\mu} \gamma_{+} (p_2)_{\nu} {\cal F}^{+}_8(s,t)
\nll[-2mm] &&
\phantom{-\frac{i}{8\ctw^2} \,\Biggl\{ \Biggl[}
    +  \gamma_{\nu} \gamma_{+} (p_1)_{\mu} {\cal F}^{+}_9(s,t)
    +  \gamma_{\nu} \gamma_{+} (p_2)_{\mu} {\cal F}^{+}_{10}(s,t) \biggr)
\ip{p_2}\varepsilon^{\sss Z}_{\nu}(p_3)  \varepsilon^{\sss Z}_{\mu}(p_4) \Biggr]    
\nll[-2mm] &&
\phantom{-\frac{i}{8\ctw^2}\,\Biggl\{ \Biggl[ }
    + \Biggl[\gamma_{+}\to\gamma_{-}, \, {\cal F}^{+}_i\to {\cal F}^{-}_{i} \Biggr] \Biggr\},
\eqa 
where
\bq
k_0 = -\frac{i g^2}{8\ctw^2}\qquad\mbox{and}\qquad\gamma_{\pm}=I\pm\gamma_{5}\,.
\eq
Furthermore,
\bq
\left( p_1+p_2 \right)^2 =-s,
\qquad
\left( p_2+p_3 \right)^2 =-t,
\qquad
\left( p_2+p_4 \right)^2 =-u.
\eq
Now we give the explicit form of the CA in the tree (Born) approximation:
\bqa
{\cal A}^{\sss Born}_{\sss ffZZ}
       &=& k_0 \,\Biggl\{
\Biggl[\vpa{f}{2}\iap{p_1}\Biggl(\frac{t+u}{t u}\,\gamma_{\mu}\sla{p_3}\gamma_{\nu}
       \gamma_{+} 
       + \frac{2}{t}\, \gamma_{\mu} \gamma_{+} (p_2)_{\nu} \     
       + \frac{2}{u}\,\biggl( \sla{p_3} \gamma_{+} \delta_{\mu\nu} 
\nll[-2mm] &&
- \gamma_{\mu} \gamma_{+} (p_1)_{\nu} + \gamma_{\nu} \gamma_{+} (p_1)_{\mu}
+ \gamma_{\nu} \gamma_{+} (p_2)_{\mu} \biggr)\Biggr)
      \ip{p_2}\varepsilon^{\sss Z}_{\nu}(p_3) \varepsilon^{\sss Z}_{\mu}(p_4) \Biggr]     
\nll[-2mm] &&
+ \Biggl[\vpa{f}{2} \to \vma{f}{2},\; \gamma_{+} \to \gamma_{-} \Biggr] \Biggr\}.
\label{BornZZ}
\eqa
Note that this is decomposed into 12 structures of 20 and is highly asymmetric in $t$ and 
$u$. This is due to our choice of the 4-momentum $p_3$ and of
the ordering of Lorentz indices $\mu$ and $\nu$ in Eq.~(\ref{BornZZ}).

Equation~\ref{BornZZ} may be parameterized by only two FFs if one introduces two 
``Born-like structures (BLS)'' given by expressions in big round brackets by means of 
eliminating the 5th structure $\sla{p_3} \gamma_{+} \delta_{\mu\nu}$ 
in favor of BLS;  to this structure and to the corresponding ${\cal F}^{\pm}_0(s,t)$
we assign the subindex ``0'':
\bqa
\sla{p_3} \gamma_{\pm}\delta_{\mu\nu}&=&-\frac{u}{2}\Bigg[{\mbox{\rm BLS}}^{\pm}_{0}
+\left(\frac{1}{t}+\frac{1}{u}\right)\gamma_{\mu}\sla{p_3}\gamma_{\nu}\gamma_{\pm}\Bigg]
- \frac{u}{t}\gamma_{\mu}\gamma_{\pm} (p_2)_{\nu}
+\gamma_{\mu}\gamma_{\pm} (p_1)_{\nu}
-\gamma_{\nu}\gamma_{\pm} (p_1)_{\mu}+\gamma_{\nu}\gamma_{\pm} (p_2)_{\mu}\;.
\eqa
%
%
Moreover, between the 20 FFs there are four identities:  
\bq
{\cal F}^{\pm}_{4}(s,t) = {\cal F}^{\pm}_{1}(s,t),\quad 
{\cal F}^{\pm}_{10}(s,t)=-{\cal F}^{\pm}_{7}(s,t).
\eq
Therefore, there are 16 independent FFs but 18 independent non-zero HAs 
for process $f_1 \bar{f}_1 \to ZZ$:
\bqa
 {\cal H}_{+-\pm\mp} &=& k^s_0 c_\pm \biggl\{
\mp 2 \vpa{\el}{2} \left(\frac{1}{t}+\frac{1}{u}\right){\cal F}^{+}_{0}(s,t)
 +\frac{s}{4}c_\mp\beta\left[2{\cal F}^{+}_{1}(s,t)-{\cal F}^{+}_{2}(s,t)-{\cal F}^{+}_{3}(s,t)
 \right]
\nll &&
 \mp 2 {\cal F}^{+}_{7}(s,t) \pm {\cal F}^{+}_{8}(s,t) \mp {\cal F}^{+}_{9}(s,t)\biggr\},
\nll
 {\cal H}_{-+\pm\mp} &=& k^s_0 c_\mp \biggl\{
 \mp 2 \vma{\el}{2} \left(\frac{1}{t}+\frac{1}{u}\right) {\cal F}^{-}_{0}(s,t)
-\frac{s}{4} c_\pm \beta\left[2{\cal F}^{-}_{1}(s,t)-{\cal F}^{-}_{2}(s,t)-{\cal F}^{-}_{3}(s,t)
\right]
\nll &&
 \mp 2 {\cal F}^{-}_{7}(s,t) \pm {\cal F}^{-}_{8}(s,t) \mp {\cal F}^{-}_{9}(s,t)\biggr\},
\nll
 {\cal H}_{+-\pm\pm} &=& k^s_0  \biggl\{
      - 2 \vpa{\el}{2} \left(\frac{\betami}{t}-\frac{\betapl}{u}\right) {\cal F}^{+}_{0}(s,t)
 -\frac{s}{4}\sin^2{\vartheta_{\sss Z}}\beta\left[2{\cal F}^{+}_{1}(s,t)-{\cal F}^{+}_{2}(s,t)
       -{\cal F}^{+}_{3}(s,t)\right]
\nll &&                 + 2 \left[
                          \beta_\mp {\cal F}^{+}_{6}(s,t)
    + \cos{{\vartheta}_{\sss Z}} {\cal F}^{+}_{7}(s,t) \right]
                      \pm  c_\mp {\cal F}^{+}_{8}(s,t)
                      \pm  c_\pm {\cal F}^{+}_{9}(s,t)\biggr\},
\nll
 {\cal H}_{-+\mp\mp} &=& k^s_0  \biggl\{
 2\vma{\el}{2}\left(\frac{\betami}{t}-\frac{\betapl}{u}\right){\cal F}^{-}_{0}(s,t)
       + \frac{s}{4}  \sin^2{{\vartheta}_{\sss Z}} \beta 
\left[ 2 {\cal F}^{-}_{1}(s,t)-{\cal F}^{-}_{2}(s,t)-{\cal F}^{-}_{3}(s,t)\right]
\nll &&
             -2\left[
                      \beta_\mp {\cal F}^{-}_{6}(s,t)
   + \cos{{\vartheta}_{\sss Z}} {\cal F}^{-}_{7}(s,t) \right]
         \mp  \frac{1}{2} c_\mp {\cal F}^{-}_{8}(s,t)
         \mp  \frac{1}{2} c_\pm {\cal F}^{-}_{9}(s,t) \biggr\},
\nll
 {\cal H}_{+-\pm0} &=& k_1^\pm  \biggl\{
 2\vpa{el}{2} 
     \left[ \frac{\betami}{t}-\frac{\betapl}{u} \pm \frac{2 \mz^2}{s}
            \left(\frac{1}{t}+\frac{1}{u}\right) \right] {\cal F}^{+}_{0}(s,t)
\nll &&
          - \frac{s}{4} c_\mp  \beta \left[
  \mp 2 \cos{{\vartheta}_{\sss Z}} {\cal F}^{+}_{1}(s,t)
	               \mp \betami {\cal F}^{+}_{2}(s,t)
	               \pm \betapl {\cal F}^{+}_{3}(s,t) \right]
\nll &&
                      \pm \betam^2 {\cal F}^{+}_{6}(s,t)
   - \left( \betapl \mp c_\mp \right) {\cal F}^{+}_{7}(s,t)
                         \mp c_\mp  {\cal F}^{+}_{8}(s,t)
                       + \betami {\cal F}^{+}_{9}(s,t) \biggr\},
\nll                           
 {\cal H}_{+-0\pm} &=& k_1^\mp  \biggl\{
2 \vpa{\el}{2} \left[\frac{\betami}{t}-\frac{\betapl}{u} 
       \mp \frac{2 \mz^2}{s} \left( \frac{1}{t}+\frac{1}{u} \right) \right] {\cal F}^{+}_{0}(s,t)
\nll &&  
    -  \frac{s}{4} c_\pm  \beta \left[
                    \pm 2\cos{{\vartheta}_{\sss Z}} {\cal F}^{+}_{1}(s,t)
	                              \pm \betami {\cal F}^{+}_{2}(s,t)
	                              \mp \betapl {\cal F}^{+}_{3}(s,t)\right]
\nll &&
                    \pm 4 \frac{\mz^2}{s} {\cal F}^{+}_{6}(s,t)     
          -\left(\betapl \pm c_\pm \right) {\cal F}^{+}_{7}(s,t)
                             - \betami {\cal F}^{+}_{8}(s,t)	   
                                \mp c_\pm {\cal F}^{+}_{9}(s,t)\biggr\},
\nll
 {\cal H}_{-+\pm0} &=& k_1^\mp  \biggl\{
 2\vma{el}{2}
\left[\frac{\betami}{t}-\frac{\betapl}{u} \mp \frac{2\mz^2}{s}
  \left(\frac{1}{t}+\frac{1}{u}\right)\right] {\cal F}^{-}_{0}(s,t)
\nll &&
      +  \frac{s}{4} c_\pm \beta  \left[ 
 \mp 2\cos{{\vartheta}_{\sss Z}} {\cal F}^{-}_{1}(s,t)
                    \mp \betami  {\cal F}^{-}_{2}(s,t)
	            \pm \betapl  {\cal F}^{-}_{3}(s,t)\right]
\nll &&
                    \mp \betap^2 {\cal F}^{-}_{6}(s,t)
- \left(\betapl \pm c_\pm \right){\cal F}^{-}_{7}(s,t)
                      \pm c_\pm  {\cal F}^{-}_{8}(s,t) 
                       + \betami {\cal F}^{-}_{9}(s,t)\biggr\},
\nll
 {\cal H}_{-+0\pm} &=& k_1^\pm  \biggl\{
 2\vma{el}{2}
\left[\frac{\betami}{t}-\frac{\betapl}{u} \pm \frac{2\mz^2}{s}
  \left(\frac{1}{t}+\frac{1}{u}\right)\right] {\cal F}^{-}_{0}(s,t)
\nll &&
      \pm \frac{s}{4} c_\mp \beta \Bigl(
2\cos{{\vartheta}_{\sss Z}}   {\cal F}^{-}_{1}(s,t)
                   + \betami  {\cal F}^{-}_{2}(s,t)
	           - \betapl  {\cal F}^{-}_{3}(s,t) \Bigr)
\nll && 
            \mp 4 \frac{\mz^2}{s} {\cal F}^{-}_{6}(s,t)
-\left(\betapl\mp c_\mp \right) {\cal F}^{-}_{7}(s,t)
	             - \betami  {\cal F}^{-}_{8}(s,t)
                     \pm c_\mp  {\cal F}^{-}_{9}(s,t) \biggr\},
\nll  
 {\cal H}_{\pm\mp00} &=& \frac{k^s_0}{2} \frac{s}{\mz^2}  \biggl\{
 \mp 2\left(\begin{array}{c}\vpa{el}{2}\\ \vma{el}{2}\end{array}\right) 
\left[\frac{\betami}{t}-\frac{\betapl}{u}
          +2\betaf\frac{\mz^2}{s} \left(\frac{1}{t}-\frac{1}{u}\right) \right] {\cal F}^{\pm}_{0}(s,t)
\nll &&
       \mp \frac{1}{4} s \beta \left[
              2 \betapl \betami  {\cal F}^{\pm}_{1}(s,t)
                   + (\betami)^2 {\cal F}^{\pm}_{2}(s,t)
	           + (\betapl)^2 {\cal F}^{\pm}_{3}(s,t)\right]
\nll &&
+ 4 \beta \frac{\mz^2}{s} 
                     {\cal F}^{\pm}_{6}(s,t)
   \pm   2 \betapl   {\cal F}^{\pm}_{7}(s,t)
   \pm \betami\left[ {\cal F}^{\pm}_{8}(s,t)-{\cal F}^{\pm}_{9}(s,t)\right] \biggr\}.
\eqa
 
Here we use the following shorthand notation:
\bqa
k^s_0    &=& \frac{k_0 s}{2}\sin{\vartheta}_{\sss Z}, \quad
k^{\pm}_1 =  \frac{k_0 s}{2}\frac{\sqrt{s}}{\sqrt{2}\mz} c_\pm\,,\quad
c_\pm     =  1 \pm \cos{\vartheta}_{\sss Z}, \qquad
\nll[1mm]
\beta_{\pm}  &=&\beta \pm 1, \qquad\quad
\beta^c_{\pm} = \beta \pm \cos{\vartheta}_{\sss Z}, \qquad
\beta         = \frac{\sqrt{\lambda(s,\mz^2,\mz^2)}}{s}\,,
\eqa
and $\vartheta_{\sss Z}$ is the CMS angle between $\vec{p}_2$ and $\vec{p}_3$.
The invariant $t$ and the cosine $\cos{\vartheta}_{\sss Z}$ are related by
\bq
 t = \mz^2-\frac{1}{2}s(1-\beta\cos{\vartheta}_{\sss Z})\,.
\eq
The number 18 is the product of 2 initial massless helicity states and
3$\times$3 states for the final $Z$ bosons.

\subsection{$f_1\bar{f}_1 \to HZ$ process}
There are six structures for the $f_1 \bar{f}_1 \to HZ$ process if the fermion
 mass is neglected
\bqa
{\cal A}_{\sss {ffHZ}} &=& k \,\Biggl\{
\Biggl[\iap{p_1} \biggl(
 \gamma_{\nu}\gamma_{+}  \vpa{f}{}  {\cal F}^{+}_0(s,t)
  +\sla{p_3} \gamma_{+} (p_1)_{\nu} {\cal F}^{+}_1(s,t)
  +\sla{p_3} \gamma_{+} (p_2)_{\nu} {\cal F}^{+}_2(s,t)\biggr) 
  \ip{p_2} \varepsilon^{\sss Z}_{\nu}(p_3) 
\Biggr]  
\nll &&
\phantom{-\frac{i g^2}{8 \ctw^2} }
+ \Biggl[\vpa{f}{} \to \vma{f}{},\;\gamma_{+}\to\gamma_{-},\;
{\cal F}^{+}_i(s,t)\to {\cal F}^{-}_{i}(s,t)
  \Biggr]\Biggr\},
\label{ffHZ-ann}
\eqa
\noindent where
\vspace*{-5mm}
\bqa
k=-\frac{ig^2}{4\ctw^2}\frac{\mz}{\mz^2-s}\,.
\eqa
The structures for the decay $H\to f_1 \bar{f}_1 Z$ may be obtained by simple replacement 
of 4-momenta $ p_1 \to -p_3,\, p_2 \to -p_4,\, p_4 \to -p_1 \, ( p_3 \to p_2)$ 
of the structures~(\ref{ffHZ-ann}).

Note, that the two terms $\gamma_{\nu}\gamma_{+}  \vpa{f}{}  {\cal F}^{\pm}_0(s,t)$
correspond to the Born level.

As far as HAs are concerned, we present them in both channels: annihilation and decay.

\subsubsection{HAs in annihilation channel $f_1\bar{f}_1\to HZ$}
There are 6 HAs in this case:
\bqa 
{\cal H}_{+-+} &=& \phantom{+}  k^s_0 \Cpl
 \Bigl\{ k^{-}_{1} \left[{\cal F}^{+}_2(s,t)-{\cal F}^{+}_1(s,t)\right] 
- 4\vpa{e}{}{\cal F}^{+}_0(s,t)\Bigr\},\nll
{\cal H}_{-++} &=& - k^s_0 \Cmi 
 \Bigl\{ k^{+}_{1} \left[{\cal F}^{-}_1(s,t)-{\cal F}^{-}_2(s,t)\right] 
+ 4\vma{e}{}{\cal F}^{-}_0(s,t)\Bigr\},\nll
{\cal H}_{+--} &=& - k^s_0 \Cmi
 \Bigl\{ k^{+}_{1} \left[{\cal F}^{+}_1(s,t)-{\cal F}^{+}_2(s,t)\right] 
+ 4\vpa{e}{}{\cal F}^{+}_0(s,t)\Bigr\},\nll
{\cal H}_{-+-} &=& \phantom{+} k^s_0 \Cpl
 \Bigl\{ k^{-}_{1} \left[{\cal F}^{-}_2(s,t)-{\cal F}^{-}_1(s,t)\right] 
- 4\vma{e}{}{\cal F}^{-}_0(s,t)\Bigr\},\nll
{\cal H}_{+-0} &=&\phantom{+} k^s_0 k_2 
  \Bigl\{{\sqrt{\lambda(s,\mz^2,\mh^2)}} 
   \left[\betapl {\cal F}^{+}_1(s,t)+\betami {\cal F}^{+}_2(s,t)\right] 
+ 4\vpa{e}{} {\cal F}^{+}_0(s,t)\Bigr\},\nll
{\cal H}_{-+0} &=&           -k^s_0 k_2 
  \Bigl\{{\sqrt{\lambda(s,\mz^2,\mh^2)}} 
   \left[\betapl {\cal F}^{-}_1(s,t)+\betami {\cal F}^{-}_2(s,t)\right] 
+ 4\vma{e}{} {\cal F}^{-}_0(s,t)\Bigr\}.
\eqa
where
\bqa
 k^s_0     &=& k_0 \frac{1}{\sqrt{2}}\frac{\sqrt{s}\mz}{s-\mz^2}\,, \qquad
 k^{\pm}_1  = \sqrt{\lambda(s,\mz^2,\mh^2)} c_\pm\,, \qquad
 k_2        = \frac{s+\mz^2-\mh^2}{\sqrt{2}\sqrt{s}\mz} \sin{\vartheta_{\sss{Z}}}, 
\nll[1mm]
 c_\pm     &=&   1 \pm \cos{\vartheta_{\sss{Z}}},\qquad
\beta^c_{\pm}= \beta\pm\cos{\vartheta_{\sss{Z}}},\qquad
\beta       =\frac{\sqrt{\lambda(s,\mz^2,\mh^2)}}{s+\mz^2-\mh^2}\,,
\nll
t &=& \mz^2-\frac{1}{2}(s+\mz^2-\mh^2)(1-\beta\cos{\vartheta_{\sss{Z}}})\,.
\eqa

\subsubsection{HAs in the decay channel $H\to f_1 \bar{f}_1 Z $\label{decaych}}
The six HAs in this case are somewhat different from the previous case:
\bqa
{\cal H}_{++-} &=& k^s_0 \Bigl\{k_1 \left[ {\cal F}^{-}_{1}(s,t)-{\cal F}^{-}_{2}(s,t)\right]
                   -4 \vma{f}{}\Cmi {\cal F}^{-}_{0}(s,t)\Bigr\}, \nll
{\cal H}_{+-+} &=& k^s_0 \Bigl\{k_1 \left[ {\cal F}^{+}_{2}(s,t)-{\cal F}^{+}_{1}(s,t)\right]
                   -4 \vpa{f}{}\Cpl {\cal F}^{+}_{0}(s,t)\Bigr\}, \nll 
{\cal H}_{-+-} &=& k^s_0 \Bigl\{k_1 \left[ {\cal F}^{-}_{1}(s,t)-{\cal F}^{-}_{2}(s,t)\right]
                   +4 \vma{f}{}\Cpl {\cal F}^{-}_{0}(s,t)\Bigr\}, \nll
{\cal H}_{--+} &=& k^s_0 \Bigl\{k_1 \left[ {\cal F}^{+}_{2}(s,t)-{\cal F}^{+}_{1}(s,t)\right]
                   +4 \vpa{f}{}\Cmi {\cal F}^{+}_{0}(s,t)\Bigr\}, \nll
{\cal H}_{0+-} &=& k^s_0 k_2 \Bigl\{ \phantom{-}
             \sqrt{\lambda{(\mh^2,\mz^2, s)}} 
                 \left[ \betapl {\cal F}^{-}_{1}(s,t)+\betami {\cal F}^{-}_{2}(s,t)\right]     
                        - 4 \vma{e}{}  {\cal F}^{-}_{0}(s,t) \Bigr\}, 
\nll
{\cal H}_{0-+} &=& k^s_0 k_2  \Bigl\{
             -\sqrt{\lambda(\mh^2,\mz^2, s)} 
                    \left[\betapl  {\cal F}^{+}_{1}(s,t)+\betami {\cal F}^{+}_{2}(s,t)\right] 
                        + 4 \vpa{e}{}  {\cal F}^{+}_{0}(s,t)\Bigr\}.
\eqa
Here,
\bqa
k^s_0 &=& k_0\frac{1}{\sqrt{2}}\frac{\sqrt{s}\mz}{\mz^2-s}\,, \qquad
k_1    =  \sqrt{\lambda(\mh^2,\mz^2, s)} \sin^2{\vartheta_f}, \qquad
k_2    =  \frac{\left(\mh^2-\mz^2-s\right) }{\sqrt{2}\sqrt{s}\mz}\sin{\vartheta_f},
\nll[2mm]
c_\pm&=&  1 \pm \cos{\vartheta_f},\qquad
\beta^c_{\pm} = \beta \pm \cos{\vartheta_f},\qquad
\beta =  \frac{\sqrt{\lambda(\mh^2,\mz^2, s)}}{\mh^2-\mz^2-s}\,.
\eqa
The number 6 is the product of 2 initial massless helicity states and 3 states
of the final $Z$ boson.

\begin{figure}[!h]
\[
\begin{picture}(132,132)(0,0)
  
 \ArrowLine(52,112)(0,112)
 \ArrowLine(52,108)(0,108)

 \Vertex(0,110){3}
 \Vertex(55,110){1}
\DashLine(0,110)(55,110){2}
 \Oval(55,110)(3,3)(0)

 \ArrowLine(0,110)(-22,132)
 \ArrowLine(0,110)( 22,88)

 \ArrowLine(58,110)(110,110)

 \ArrowArc(0,110)(22,0,135)
 \Text(15,127)[lb]{$\vartheta_f$} 

 \Text(25,116)[lb]{$p_3$} 
 \Text(25,99)[lb]{$p_4$}
 
 \Text(80,99)[lb]{$p_1$} 
 \Text(55,99)[lb]{$p_2$} 
 
 \Text(-25,116)[lb]{$p_3$}
 \Text(-4,93)[lb]{$p_4$}

\end{picture}
\]
\vspace*{-40mm}
\caption[$H\to f_1\bar{f}_1 Z$ decay kinematics]
        {$H\to f_1\bar{f}_1 Z$ decay kinematics.}
\label{Blevel}
\end{figure}
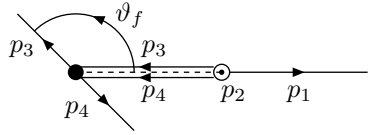

Furthermore, $s=M^2_{f_1 \bar{f}_1}$
is the invariant mass of the two fermions $f$, varying in the 
limits $4 m^2_{f} \leq s \leq (\mh-\mz)^2$;
and $t$ is another independent kinematical variable, depending on $s$ and an angle
$\vartheta_f$, varying in the limits $0\leq\vartheta_f\leq\pi$ 
\bq
t=\mz^2+\frac{1}{2}\left[\mh^2-\mz^2-s-\sqrt{\lambda(\mh^2,\mz^2, s)}\cos{\vartheta_f}\right].
\eq
The kinematical diagram of the process is shown in Fig.~\ref{Blevel}.

The Higgs boson with momentum $p_2$ at rest, decays back-to-back into a $Z$
 boson with momentum 
$p_1$ and a fermionic compound with 4-momentum $p_3+p_4$ and invariant mass $s$.
This compound decays in its own rest frame into two back-to-back fermions
 with $\vartheta_f$
being the angle between $p_3$ in the compound rest frame and the direction
of flight of the $Z$ boson in the $H$ boson rest frame.

\clearpage

\section{Precomputation news \label{precomputation}}
\begin{figure}[!t]
\[
\begin{picture}(125,90)(0,0)
  \Vertex(100,43){12.5}
  \Vertex(25,43){2.5}
  \Photon(125,86)(100,43){2}{10}
  \Photon(100,43)(125,0){2}{10}
  \Photon(25,43)(100,43){3}{10}
  \ArrowLine(0,0)(25,43)
  \ArrowLine(25,43)(0,86)
 \Text(-15,90)[lb]{fd=typeID}
 \Text(-15,-13)[lb]{fu=typeIU}
 \Text(110,90)[lb]{vu=typeFU}
 \Text(110,-13)[lb]{vd=typeFD}

 \ArrowLine(3,0)(13,16)
 \ArrowLine(3,86)(13,70)

 \Text(62.5,48)[cb]{$B$}
 \Text(11,76)[lb]{$p_1$}
 \Text(10,3)[lb]{$p_2$}
 \Text(127,76)[lb]{$b$}
 \Text(127,3)[lb]{$b$}

 \ArrowLine(122,86)(112,70)
 \ArrowLine(121,-2)(111,14)

 \Text(105,76)[lb]{$p_4$}
 \Text(104,3)[lb]{$p_3$}

\end{picture}
\]
\vspace*{-2mm}
\caption[``Right'' $bbb$ vertex in $ffbb$ processes]
        {``Right'' $bbb$ vertex in $ffbb$ processes.}
\label{Vertbbb}
\end{figure}
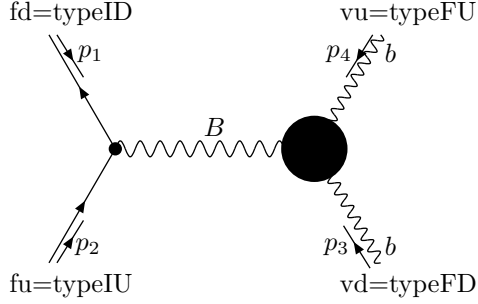

The ``Precomputation'' tree of {\tt version 1.10} 
is shown in Fig.~\ref{PrecEW11} with all modified sub-menus open, 
and all sub-menus closed which were not changed compared to 
{\tt version 1.00}. In this section we briefly discuss what every new module does.
\begin{floatingfigure}{73mm}
\hspace*{-6mm}
\includegraphics[width=73mm,height=164mm]{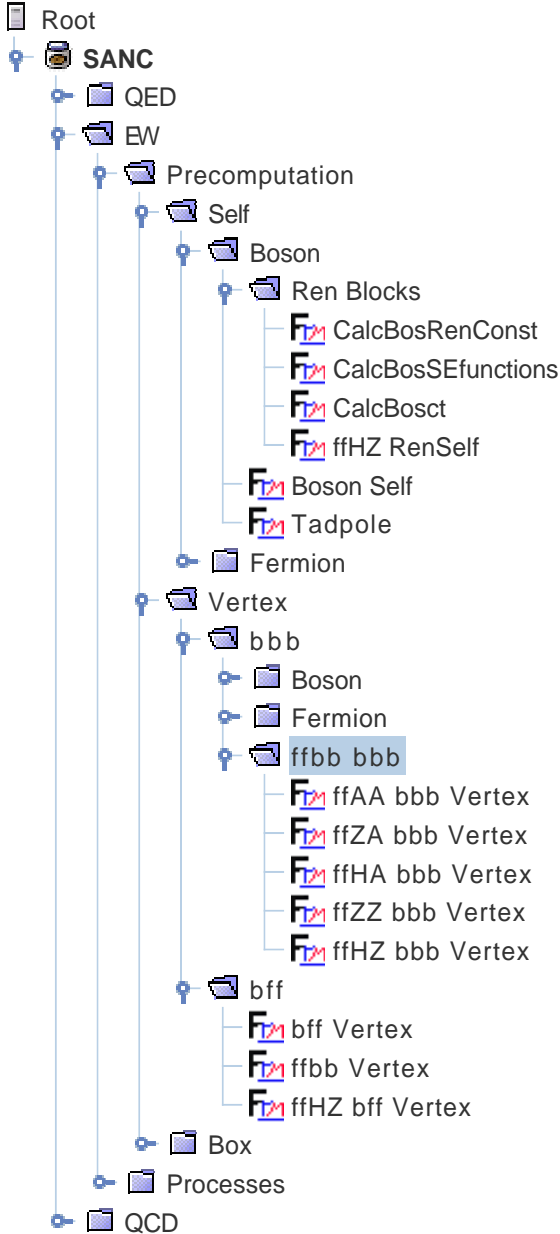}
\caption[New EW precomputation modules]
        {New EW precomputation modules.}\label{PrecEW11}
\end{floatingfigure}
First, we added a new folder accessible via menu sequence
{\bf EW $\to$ Precomputation $\to$ Vertex $\to$ bbb $\to$ ffbb bbb} with five modules
{\bf ffXX bbb Vertex, XX=AA, ZA, HA, ZZ, HZ} which compute three boson vertices of
 four topologies
(see Fig.~11 of Ref.~\cite{Andonov:2004hi}) to the corresponding $ffbb$ processes, 
Fig.~\ref{Vertbbb}. The results of their calculations are saved to {\sf ffXX*.sav} files 
to be loaded by corresponding modules computing FFs via chains
{\bf EW $\to$ Processes $\to$ 4-legs $\to$ 2f2b $\to$ Neutral Current} for
 the $ffXX\to 0$
processes.  

These three boson diagrams contain both {\em bosonic} and {\em fermionic} components.
The latter are precomputed by the modules {\bf bbb Vertex} in {\bf Boson} and
 {\bf Fer\-mion} 
folders of the same level on the tree. Five modules of {\bf ffbb bbb} folder
 load them and
then apply tedious calculations involving in some cases the Schouten identity. 
There are many peculiarities in these calculations, forcing us to have an
 individual module
for each $ffbb$ process. Note also that if the corresponding process has a Born-level
$s$ channel exchange as in Fig.~\ref{Borns}, then the contribution of one-loop vertices 
is supplemented by the relevant counterterm cross. 

In the modules under discussion a summation over the exchanged boson $B$ is performed. 
In general, four neutral bosons $B=\gamma,Z,\phi^{0}$ and $H$ can
 contribute if the fermion
mass is not neglected, otherwise, only $\gamma$ and $Z$ contribute.

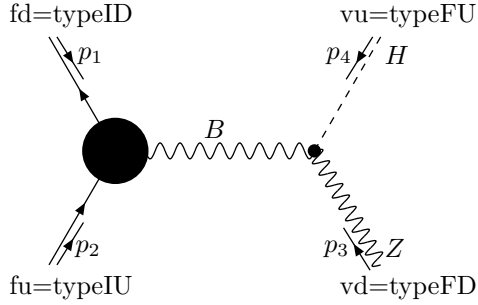
\begin{figure}[!h]
\[
\begin{picture}(125,90)(0,0)
  \Vertex(100,43){2.5}
  \DashLine(125,86)(100,43){3}
  \Photon(100,43)(125,0){3}{10}
  \Photon(25,43)(100,43){3}{10}
  \Vertex(25,43){12.5}
  \ArrowLine(0,0)(25,43)
  \ArrowLine(25,43)(0,86)
 \Text(62.5,48)[cb]{$B$}
 \Text(-15,90)[lb]{fd=typeID}
 \Text(-15,-13)[lb]{fu=typeIU}
 \Text(110,90)[lb]{vu=typeFU}
 \Text(110,-13)[lb]{vd=typeFD}

 \ArrowLine(3,0)(13,16)
 \ArrowLine(3,86)(13,70)

 \Text(11,76)[lb]{$p_1$}
 \Text(10,3)[lb]{$p_2$}
 \Text(127,76)[lb]{$H$}
 \Text(127,3)[lb]{$Z$}

 \ArrowLine(122,86)(112,70)
 \ArrowLine(121,-2)(111,14)

 \Text(105,76)[lb]{$p_4$}
 \Text(104,3)[lb]{$p_3$}

\end{picture}
\]
\vspace*{-2mm}
\caption[``Left'' $bff$ vertex in $ffbb$ processes]
        {``Left'' $bff$ vertex in $ffbb$ processes.}
\label{Vertbbf}
\end{figure}

For the processes $ffAA\to 0$, $ffZA\to 0$, $ffHA\to 0$ the
 ``left'' $bff$ vertex diagram,
shown in Fig.~\ref{Vertbbf} does not contribute, since in these cases the ``right'' 
vertex does not exist at the tree level. In general, and this is indeed the case for 
the processes $ffZZ\to 0$ and $ffHZ\to 0$, the ``right'' vertex exists at the tree level
for $B=Z,\phi^{0}$, therefore, the dressed ``left'' vertex has to be added to the 
precomputation tree. Note that it does not contribute for massless fermions if $B=\phi^{0}$.
For $ffHZ\to 0$ this vertex is accessible via menu sequence
{\bf EW $\to$ Precomputation $\to$ Vertex $\to$ bff $\to$ ffHZ bff Vertex}.
For the $ffZZ\to 0$ process, only $B=H$ contributes, but then, for the massless $f$,
the dressed ``left'' vertex vanishes. This is why we do not add the corresponding module in
the latter case.

\clearpage

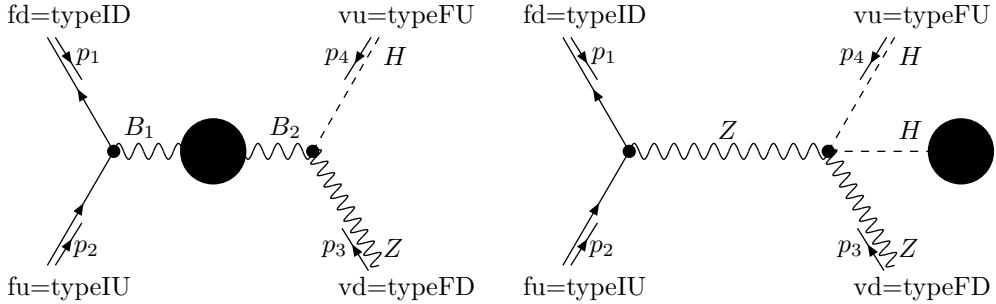
\begin{figure}[!h]
\[
\begin{array}{cc}
\begin{picture}(125,86)(0,0)
  \Vertex(100,43){2.5}
  \DashLine(125,86)(100,43){3}
  \Photon(100,43)(125,0){3}{10}
  \Photon(25,43)(100,43){3}{10}
  \Vertex(25,43){2.5}
  \Vertex(62.5,43){12.5}
  \ArrowLine(0,0)(25,43)
  \ArrowLine(25,43)(0,86)
 \Text(-15,90)[lb]{fd=typeID}
 \Text(-15,-13)[lb]{fu=typeIU}
 \Text(110,90)[lb]{vu=typeFU}
 \Text(110,-13)[lb]{vd=typeFD}

 \ArrowLine(3,0)(13,16)
 \ArrowLine(3,86)(13,70)

 \Text(35,48)[cb]{$B_1$}
 \Text(90,48)[cb]{$B_2$}
 \Text(11,76)[lb]{$p_1$}
 \Text(10,3)[lb]{$p_2$}
 \Text(127,76)[lb]{$H$}
 \Text(127,3)[lb]{$Z$}
 \ArrowLine(122,86)(112,70)
 \ArrowLine(121,-2)(111,14)

 \Text(105,76)[lb]{$p_4$}
 \Text(104,3)[lb]{$p_3$}

\end{picture}
&\qquad\qquad\qquad
\begin{picture}(125,86)(0,0)
  \Vertex(100,43){2.5}
  \DashLine(125,86)(100,43){3}
  \Photon(100,43)(125,0){3}{10}
  \Photon(25,43)(100,43){3}{10}
  \Vertex(25,43){2.5}
  \ArrowLine(0,0)(25,43)
  \ArrowLine(25,43)(0,86)
  \DashLine(100,43)(140,43){3}
  \Vertex(150,43){12.5}
 \Text(-15,90)[lb]{fd=typeID}
 \Text(-15,-13)[lb]{fu=typeIU}
 \Text(110,90)[lb]{vu=typeFU}
 \Text(110,-13)[lb]{vd=typeFD}

 \ArrowLine(3,0)(13,16)
 \ArrowLine(3,86)(13,70)

 \Text(62.5,48)[cb]{$Z$}
 \Text(127.,48)[lb]{$H$}
 \Text(11,76)[lb]{$p_1$}
 \Text(10,3)[lb]{$p_2$}
 \Text(127,76)[lb]{$H$}
 \Text(127,3)[lb]{$Z$}

 \ArrowLine(122,86)(112,70)
 \ArrowLine(121,-2)(111,14)

 \Text(105,76)[lb]{$p_4$}
 \Text(104,3)[lb]{$p_3$}

\end{picture}
\end{array}
\]
\vspace*{-2mm}
\caption[Self energy $ffHZ$ diagrams]
        {Self energy $ffHZ$ diagrams.}
\label{Selfsch}
\end{figure}

The presence of an $s$ channel tree level diagram in the process $f_1\bar{f}_1 HZ\to 0$ 
(Fig.~\ref{Borns}) forces us to take into account two more self energy diagrams,
Fig.~\ref{Selfsch}.

The first one is accessible via menu chain:
{\bf Self $\to$ Boson $\to$ Ren Blocks $\to$ ffHZ Ren Self}. Here $B_1=\gamma,Z,\phi^{0}$
and $B_2=Z,\phi^{0}$; again $\phi^{0}$'s do not contribute for massless fermions.
As will be explained in the next section, the second diagram
is better to be combined with the ``right'' vertex, Fig.~\ref{Vertbbb}.

Note that nothing is changed, compared to $ffXA$ processes,
as far as boxes are concerned. So, the {\bf Box} sub-menu is as 
{\tt in version 1.00}~\cite{Andonov:2004hi}.

The new modules contain calls to several new {\it intrinsic} procedures
which will be described elsewhere. 

\clearpage 
\section{Renormalizaton for $ffHZ\to 0$ process\label{renormalization}}
In this section we describe how to use the FORM~\cite{Vermaseren:2000nd} module which computes 
FFs for the process $f f HZ\to 0$. This description is supposed to help a user to understand 
the other modules computing FFs for any $ffbb\to 0$ process.

First of all, to use our basic declaration and notation we begin the file with
\begin{verbatim}
#include Declar.h
#call Globals()
\end{verbatim}
and define {\sf types} of external particles, see Figs.~\ref{Borntu} and~\ref{Borns}. 
\begin{verbatim}
#ifdef `typeIU';  *  `fu'
#ifdef `typeID';  *  `fd'
#ifdef `typeFU';  *  `vu'
#ifdef `typeFD';  *  `vd'
#define typeIDp "{2*(`typeID'%2)-1+`typeID'}";  * `fdp'
\end{verbatim}

Secondly, we fix four main steering flags to define the calculation Al scheme:
\begin{enumerate} 
 \item{define {\sf xi}:} {\sf xi = 0} to test gauge invariance
 in $R_\xi$, or {\sf xi = 1} 
to work in $\xi=1$ gauge;

 \item{define {\sf on}:} 
   {\sf on = 0} external photons are off mass-shell,
or {\sf on = 1} photons are on mass-shell;

\item{define {\sf mf}:}
   {\sf mf = 0} zero external fermion mass ({\it i.e.} {\sf pm(`fd')=0}),
or {\sf mf = 1} it is not zeroed;

\item{define {\sf mp}:}
  {\sf mp = 0} zero mass of the weak isospin partner of the fermion $f$, {\sf pm(`fdp')=0}, 
or {\sf mp = 1} it is not zeroed.

\end{enumerate}

Actually, for the process under consideration, $f_1\bar{f}_1 HZ\to 0$,
 only the $\xi$ and {\sf mp} definitions
are meaningful since there are no external photons and we ignore
the masses of external fermions throughout the calculations.
But the mass of a weak isospin partner of an external fermion that appears
in the internal loop may be kept nonzero.

The ideology of building blocks (BB) is the key element for {\tt SANC} development.
The information about the main precomputed BB is stored in basic {\bf *.sav} files (BSF).
Note that for 4-particle $ffbb$ processes all the BBs are 4-legs by construction.
 This trick
will greatly simplify the procedure of projection of the covariant amplitude
onto an independent basis of structures.

Moreover, for the future development it is necessary to upgrade the database of
the collected information area {\tt SANC}: fields of program modules, fields of
 procedures and  
the bank of BSF.

Any module computing FFs starts from loading of the calculated BB from the bank of 
BSF. These BSF contain the precomputed objects: self energies,
 vertices and boxes typically
with off-shell bosons.

They are precomputed not only to accelerate the calculations. Although all BSF may be,
in principle, precomputed online, we remind that in some cases the CPU time for
calculating off-shell boxes in $R_\xi$ gauge takes many hours, see section 3.4 of 
Ref.~\cite{Andonov:2004hi}. In such cases precomputation is strictly prohibited
and the user must use already precomputed BSFs.

We recall also that our precomputation procedure has indeed several levels;
 in the modules
computing FFs we tend to use the results of the last level which contains already
renormalized BBs: propagators and vertices, {\it i.e.} taking into account relevant 
{\em counterterms} and {\em special vertices},~\cite{Bardin:1999ak}. 
However, they are full of residual UV poles and $\xi$ dependent terms,
 which cancel in the sum 
for a one-loop CA of a physical process. This is why we still use to word 
``renormalization'' in connection with modules computing FFs rather
 than a simple ``summation''.

Typically, the loading of BSFs is organized in several steps.
Let us consider the example of
\linebreak {\bf H $\to$ f1 f1 Z (FF)} module, see the tree in Fig.~\ref{PrecEW11}.

\begin{itemize}
\item{\underline{step self}}

Here we manipulate objects from BSF {\sf ffHZSelfschxi`xi'`fu'`fd'`vu'`vd'.sav}

We extract from its volume the BB of bosonic self energy in the s-channel, 
{\sf BSEsch`fu'`fd'`vu'`vd'} see left diagram in Fig.~\ref{Selfsch}. 

\item{\underline{step vertex}}

Moving further over the renormalization procedure at this step we load three BSFs:\\
{\sf ffHZVertbffxi`xi'`fu'`fd'`vu'`vd'.sav};\\
{\sf ffHZVertbbbxi`xi'`fu'`fd'`vu'`vd'.sav};\\
{\sf ffbbVertxi`xi'on`on'mf`mf'mp`mp'`fu'`fd'`vu'`vd'.sav}.

From these BSFs we extract various types of vertices correspondingly:\\
--- {\sf VertBff`i'`fu'`fd'`vu'`vd'} with {\sf i=1,2,3,4} standing for $\xi_{\sss A}$, 
$\xi_{\sss Z}$, $\xi_{\sss W}$, and {\sf no} $\xi$ {\em vertex clusters}
 originating from 
the diagram of Fig.~\ref{Vertbbf};\\
--- {\sf Vertbbbbos`fu'`fd'`vu'`vd'} and {\sf Vertbbbfer`fu'`fd'`vu'`vd'} --- the
{\em bosonic} and {\em fermionic} components of three-boson vertices
 shown in the diagram 
Fig.~\ref{Vertbbb}, where the former contains counterterms, the special
 vertex and the right 
diagram of Fig.~\ref{Selfsch}. These tadpoles cancel
the $\xi_{\sss Z}$ dependence of 
the three-boson vertices, giving an opportunity to assign this
 contribution to {\sf i=3}.
Finally, the fermionic component should be naturally assigned to {\sf i=4};\\
--- abelian {Vert`I'`i'} and non-abelian {vert`I'`i'} vertex clusters
 in $t$ and $u$ channels
{\sf I=t,u} with cluster index {\sf i=1,2,3,4} and {\sf k=22,33,24,42,44},
 see section 3.4.2
of Ref.~\cite{Andonov:2004hi} for a description of the latter.

\item{\underline{step boxes}}

Here the most complex building blocks --- off-shell boxes are loaded from four BSFs:\\
{\sf ffbb3T1xi`xi'on`on'mf`mf'mp`mp'`fu'`fd'`vd'`vu'.sav};\\
{\sf ffbb3T3xi`xi'on`on'mf`mf'mp`mp'`fu'`fd'`vu'`vd'.sav};\\
{\sf ffbb22T5xi`xi'on`on'mf`mf'mp`mp'`vd'`fd'`vu'`fu'.sav};\\
{\sf ffbb33T5xi`xi'on`on'mf`mf'mp`mp'`vd'`fd'`vu'`fu'.sav}.
\vskip 2pt

They contain precomputed boxes of topology T1 from the expression
{\sf S3T1`xi'`on'`mf'`mp'`fu'`fd'} {\sf`vd'`vu'},
the boxes of topology T3 from the expression
{\sf S3T3`xi'`on'`mf'`mp'`fu'`fd'`vu'`vd'},
and of topology T5 with cluster index $k1=2,3$ , {\it i.e.} with virtual
$Z$ and $W$ bosons from the expression
{\sf S`k1'`k1'T5`xi'`on'`mf'`mp'`vd'`fd'`vu'`fu'}, see section 3.4.2
 of Ref.~\cite{Andonov:2004hi}.

Only those box topologies and clusters are loaded which give a non-zero contribution 
for $m_{f}=0$.

\item{\underline{step Sum}}

Finally, we sum all contributions. 
Four expressions, {\sf Sum`i'}, corresponding to cluster index {\sf i=1,2,3,4}
are being constructed here.
The first three of them may carry only one gauge parameter each,
the latter carries none.
\end{itemize}

After construction of four {\sf Sum`i'}, the module continues with various kinds of 
transformations (in particular, involving an algebra of Gram determinants)
which prove the cancellation of gauge parameter 
dependences in first three {\sf Sum1,2,3} and the cancellation of the
residual UV poles between FFs with cluster indices {\sf i=3} and {\sf i=4}.

After the comment \underline{\tt Preparing Structures for obtaining FFs:} the
 six basis elements
of the CA, Eq~(\ref{ffHZ-ann}), are created and the 6$\times 4$ FFs are
 projected out of this CA.
 
The final step is formatting of the BSF {\sf FFf1f1HZ.sav} with $24$
FFs --- \underline{FFgp`k'`i'} and \underline{FFgm`k'`i'} ($k=0,1,2$, $i=1,4$) ---
for subsequent processing by {\sf s2n.f} software.  


\section{Bremsstrahlung in $f_1\bar{f}_1 HZ\to 0$ processes\label{brems}}
In this section we present the list of short final results for the contribution of 
accompanying bremsstrahlung processes.
\subsection{Bremsstrahlung in $f_1\bar{f}_1\to HZ$ annihilation channel}
The tree level diagram of this channel is shown in Fig.~\ref{Borns}. 
The corresponding total Born cross section reads:
\bqa
\sigma^{\rm Born}=
\frac{G_{\sss F}^2 (v_f^2+a_f^2) }{12\pi} 
         \frac{\mz^4 \sqrtL{s}{\mz^2}{\mh^2}}{(\mz^2 - s)^2 + \mz^2 \Gamma^2_{\sss Z}}
\left[  \frac{1}{2} 
 + \frac{1}{s} \left(5\mz^2-\mh^2\right)+\frac{1}{2 s^2}\left(\mh^2-\mz^2\right)^2\right].
\label{Bornann} 
\eqa

There are only two initial state (ISR) bremsstrahlung diagrams:

\begin{figure}[!h]
\[
\begin{array}{cc}
\begin{picture}(125,86)(0,0)
 \Vertex(100,43){2.5}
 \DashLine(125,86)(100,43){3}
 \Photon(100,43)(125,0){3}{10}
 \Photon(25,43)(100,43){3}{10}
 \Vertex(25,43){2.5}
 \ArrowLine(0,0)(25,43)
 \ArrowLine(25,43)(0,86)
 \ArrowLine(3,0)(13,16)
 \ArrowLine(3,86)(13,70)
 \Text(11,76)[lb]{$p_1$}
 \Text(10,3)[lb]{$p_2$}
 \Text(40,66)[cb]{$p_5$}
 \ArrowLine(30,64)(50,64)
 \Photon(15,60)(50,60){2}{6}
 \Text(127,76)[lb]{$H$}
 \ArrowLine(112,70)(122,86)
 \Text(105,76)[lb]{$p_4$}
 \Text(127,3)[lb]{$Z$} 
 \ArrowLine(111,14)(121,-2)
 \Text(104,3)[lb]{$p_3$}
\end{picture}
&\qquad\qquad
\begin{picture}(125,86)(0,0)
 \Vertex(100,43){2.5}
 \DashLine(125,86)(100,43){3}
 \Photon(100,43)(125,0){3}{10}
 \Photon(25,43)(100,43){3}{10}
 \Vertex(25,43){2.5}
 \ArrowLine(0,0)(25,43)
 \ArrowLine(25,43)(0,86)
 \ArrowLine(3,0)(13,16)
 \ArrowLine(3,86)(13,70)
 \Text(11,76)[lb]{$p_1$}
 \Text(10,3)[lb]{$p_2$}
 \Text(40,18)[ct]{$p_5$}
 \ArrowLine(30,21)(50,21)
 \Photon(14.5,25)(50,25){2}{6}
 \Text(127,76)[lb]{$H$}
 \ArrowLine(112,70)(122,86)
 \Text(105,76)[lb]{$p_4$}
 \Text(127,3)[lb]{$Z$} 
 \ArrowLine(111,14)(121,-2)
 \Text(104,3)[lb]{$p_3$}

\end{picture}
\end{array}
\]
\vspace*{-2mm}
\caption[Bremsstrahlung diagrams in annihilation channel]
        {Bremsstrahlung diagrams in annihilation channel}
\label{Brems}
\end{figure}
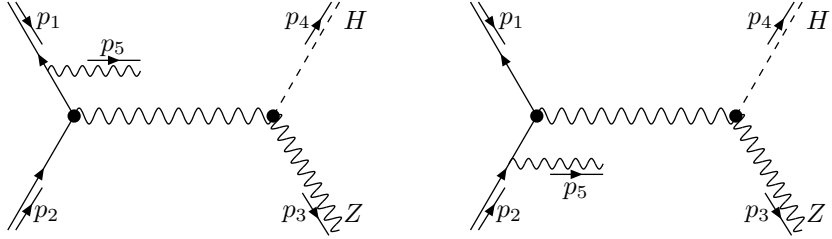

QED corrections due to virtual and soft photons are proportional to the
 Born cross section:
\bqa
\sigma^{\rm Virt}={\sigma^{\rm Born}}
 \frac{\alpha}{\pi} \qel^2
       \left\{ \frac{1}{2} \left[\ln\left(\frac{s}{\mf^2}\right)-1\right]^2 
             + \left[\frac{3}{2} - \ln\left(\frac{s}{\lambda^2}\right)\right]
               \left[\ln\left(\frac{s}{\mf^2}\right)-1\right]
              - 1  + 4 \Litwo(1) \right\},
\label{Virtann} 
\eqa
\bqa
\sigma^{\rm Soft}={\sigma^{\rm Born}}
\frac{\alpha}{\pi} \qel^2
       \left\{ -\frac{1}{2} \left[\ln\left(\frac{s}{\mf^2}\right)-1\right]^2 
           + \ln\left(\frac{4 \bar\omega^2}{\lambda^2}\right) 
             \left[\ln\left(\frac{s}{\mf^2}\right)-1\right] 
           + \frac{1}{2} - 2 \Litwo(1) \right\},
\label{Softann} 
\eqa
with infrared divergence $\ln{\lambda^2}$ being canceled 
out\footnote{Note, that the ``Soft'' contribution to the process $f_1\bar{f}_1\to ZZ$
is also described by~Eq.(\ref{Softann}).}.
The hard photon contribution of the differential cross section in
$s'=-(p_3+p_4)^2$ has the following factorization property:
\bqa
\frac{d\sigma^{\rm Hard}}{d s'}=
\frac{\alpha}{\pi} \qel^2 \frac{s^2+{s'}^2}{s^2 (s-s')}
               \left[\ln\left(\frac{s}{\mf^2}\right)-1\right]
\sigma^{\rm Born}(s')\,.
\label{Hardann}
\eqa
It may be integrated over $s'$ leading to a rather compact expression for $\sigma^{\rm{tot}}$
(see relevant module at {\tt SANC} tree), yielding a $\ln{\bar\omega}$ which cancels 
against corresponding term in Eq.~(\ref{Softann}).


\subsection{Bremsstrahlung in $H\to f_1\bar{f}_1 Z$ decay channel}
Here we consider the decay channel $H\to f_1 \bar{f}_1 Z$.
We begin with the tree level diagram, Fig.~\ref{tree}:

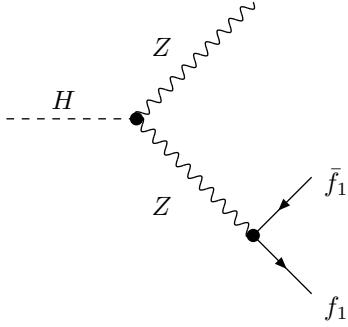
\begin{figure}[!h]
\vspace*{-10mm}
\[
\begin{array}{cc}
\begin{picture}(132,132)(0,0)
 \DashLine(-5,66)(44,66){3}
 \Vertex(44,66){2.5}
 \Text(12, 70)[lb]{$H$}
\Photon(44,66)(88,110){2}{10}
\Photon(44,66)(88,22){2}{10}
 \Vertex(88,22){2.5}
 \Text(50,90)[lb]{$Z$}
\ArrowLine(110,44)(88,22)
\ArrowLine(88,22)(110,0)
 \Text(115,37)[lb]{$\bar{f}_1$}
 \Text(115,-10)[lb]{$f_{1}$}
 \Text( 50,30)[lb]{$Z$}
\end{picture}
\end{array}
\]
\vspace*{-3mm}
\caption[The $H\to f_1\bar{f}_1 Z$ decay tree level diagram\label{tree}]
        {The $H\to f_1\bar{f}_1 Z$ decay tree level diagram\label{tree}.} 
\end{figure}

The corresponding tree level double differential width, 
depending on two kinematical variables $s,\;\vartheta_f$ discussed in
section~\ref{decaych} and with kinematics shown in Fig.~\ref{Blevel}, reads
\bqa
\frac{d^2\Gamma^{\rm Born}}{ds\,d\cos\vartheta_f} &=& k_{\sss B}
\Biggl\{   \Biggl(   
\left(v_f^2+a_f^2\right) 
\left[\sin^2\theta_f\left(1-4\frac{\mf^2}{s}\right)+4\frac{\mf^2}{s}\right]-8 a_f^2\frac{\mf^2}
{\mz^2} \Biggr)
\nll &&
 \left[ \left( 1-\frac{s}{\mh^2} \right)^2
     - 2 \frac{s \mz^2}{\mh^4}-1+\left(1-\frac{\mz^2}{\mh^2}\right)^2 \right]
+8 \left( v_f^2+a_f^2 \right) \frac{s \mz^2}{\mh^4}\left(1+2\frac{\mf^2}{s}\right)
\nll &&
+4  a_f^2 \frac{\mf^2}{\mz^2}
  \left[  \frac{s \mz^2}{\mh^4}\left(1-\frac{s}{\mz^2}\right)^2
   + \frac{s}{\mh^2}\left( - 2\frac{s}{\mz^2} - 2 + \frac{\mh^2}{\mz^2}\right)
         - 12\frac{\mz^4}{\mh^4} \right]  \Biggr\},                                              
\nll &&
\mbox{where} \hspace*{4mm} k_{\sss B}=\frac{1}{128}\frac{G^2_{\sss F}}{\pi^3}
\frac{\sqrtL{\mh^2}{\mz^2}{s}\mz^4\mh}{\left|\mz^2-i\mz\Gamma_{\sss Z} - s\right|^2}\,.
\eqa

There are only two final state bremsstrahlung diagrams, Fig.~\ref{Brem}:

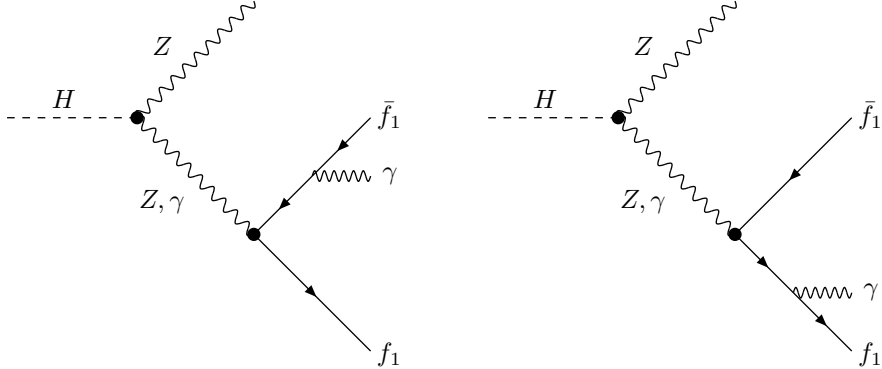
\begin{figure}[!h]
\vspace*{-3mm}
\[
\begin{array}{cc}
\begin{picture}(132,132)(0,-25)
{\DashLine(-5,66)(44,66){3}
 \Vertex(44,66){2.5}
 \Text(12, 70)[lb]{$H$}
\Photon(44,66)(88,110){2}{10}
\Photon(44,66)(88,22){2}{10}
 \Vertex(88,22){2.5}
 \Text(50,90)[lb]{$Z$}
\ArrowLine(110,44)(88,22)
\ArrowLine(132,66)(110,44)
\ArrowLine(88,22)(132,-22)
 \Text(135,62)[lb]{$\bar{f}_1$}
 \Text(135,-28)[lb]{$f_{1}$}
 \Text(45,30)[lb]{$Z,\gamma$}
  \Photon(110,44)(132,44){2}{6}
  \Text(137,41)[lb]{$\gamma$}
}
\end{picture}
&\qquad\qquad
\begin{picture}(132,132)(0,-25)
{\DashLine(-5,66)(44,66){3}
 \Vertex(44,66){2.5}
 \Text(12, 70)[lb]{$H$}
\Photon(44,66)(88,110){2}{10}
\Photon(44,66)(88,22){2}{10}
 \Vertex(88,22){2.5}
 \Text(50,90)[lb]{$Z$}
\ArrowLine(132,66)(88,22)
\ArrowLine(88,22)(110,0)
\ArrowLine(110,0)(132,-22)
  \Text(135,62)[lb]{$\bar{f}_1$}
  \Text(135,-28)[lb]{$f_{1}$}
  \Text(45,30)[lb]{$Z,\gamma$}
\Photon(132,0)(110,0){2}{6}
\Text(137,-3)[lb]{$\gamma$}
}
\end{picture}
\end{array}
\]
\vspace*{-5mm}
\caption[The $H\to f_1 \bar{f}_1 Z$ decay, bremsstrahlung]
        {The $H\to f_1 \bar{f}_1 Z$ decay, bremsstrahlung.\label{Brem}}
\end{figure}

\newpage

The fully differential phase space is characterized by five kinematical variables which 
we choose as follows:
\bqa
d\Phi^{(3)}=\frac{ds}{2\pi}\frac{d\tau}{2\pi}\,\Phi^{(2)}_1\,d\Phi^{(2)}_2\,d\Phi^{(2)}_3\,,
\eqa
where $\tau=-(p_4+p_5)^2$ is the lepton--photon invariant mass.

\begin{figure}[!h]
\[
\begin{picture}(132,132)(0,0)
 \ArrowLine(52,113.5)(0,113.5)
 \ArrowLine(52,106.5)(0,106.5)
 \ArrowLine(52,110)(0,110)
 \Vertex(0,110){5}
 \Vertex(55,110){1}
\DashLine(0,110)(55,110){2}
 \Oval(55,110)(5,5)(0)

 \ArrowLine(0,110)(-35,132)

 \Line(3,110)(37,90)
 \ArrowLine(17,101.5)(18,101)
 \ArrowLine(-3,110)(35,86.5)

 \ArrowArc(0,110)(23,0,149)
 \Text(15,127)[lb]{$\vartheta_f=\vartheta_3$} 

 \Text(23,116)[lb]{$p_{3,4,5}$} 
 
 \Text(80,97)[lb]{$p_1$} 
 \Text(55,97)[lb]{$p_2$} 
 
 \Text(-20,111)[lb]{$p_3$}
 \Text(-4,93)[lb]{$p_{4,5}$}

\ArrowLine(58,110)(110,110)
\SetScale{0.5}
\Line(216,90)(216,203)
\SetScale{1}
\ArrowLine(108,101)(108,103)
 \Text(140,29)[lb]{$z$}
\SetScale{0.5}
\Line(216,90)(288,46)
\Line(216,90)(144,132) 
\Line(216,90)(272,132)
\SetScale{1}
\DashLine(0,110)(72,66){3}
  \ArrowLine(72,66)(71,66.5)
 \Text(71,29)[lb]{$x$}
\SetScale{0.5}
\Line(216,90)(148,46)
\SetScale{1}
\ArrowLine(74.5,23)(73,22)
\DashLine(0,110)(72,66){3}
\ArrowLine(108,45)(86,20)
\ArrowLine(108,45)(130,70)
\ArrowArc(108,45)(27,48,-210)     
 \Text(112,60)[lb]{$p_4$}
 \Text(112,96)[lb]{$y$}
 \Text(100,25)[lb]{$p_5$}
 \Text(80,70)[lb]{$\vartheta_4$}
\Vertex(95,90){1}
\DashLine(130,70)(95,90){1}
\DashLine(95,90)(95,38){1}
\DashLine(95,90)(108,45){1}
\ArrowArc(108,45)(-24,285,34)

\DashLine(95,90)(108,98){1}

 \Text(72,44)[lb]{$\varphi_4$}
\end{picture}
\]
\vspace*{-15mm}
\caption[Bremsstrahlung kinematics]
        {Bremsstrahlung kinematics.}
\label{cascade}
\end{figure}
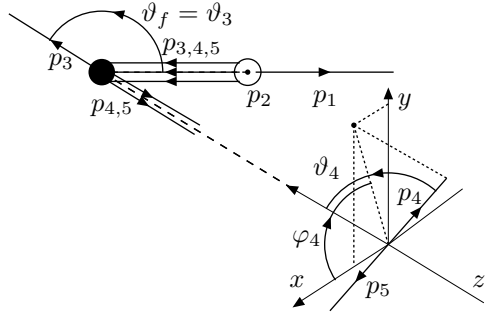

The 3-step kinematical cascade develops as a sequence of three 2-body decays shown in 
Fig.~\ref{cascade},

\noindent
with three corresponding two body phase spaces:
\bqa
 \Phi^{(2)}_1 &=& \frac{1}{8\pi}\frac{\sqrtL{\mh^2}{\mz^2}{s}}{s}\,,
\nll
d\Phi^{(2)}_2 &=& \frac{1}{8\pi}\frac{\sqrtL{s}{\tau}{m^2_{f}}}{s}\,
                                \frac{1}{2}\,d\cos\vartheta_3\,,
\nll
d\Phi^{(2)}_3 &=& \frac{1}{8\pi}\frac{\sqrtL{\tau}{m^2_{f}}{0}}{\tau}\,
                                \frac{1}{2}\,d\cos\vartheta_4\,\frac{1}{2\pi}d\varphi_4\,.
\eqa
The gauge invariant QED part of the complete one-loop EW correction is subdivided
into {\it virtual, soft} and {\it hard} photon contributions. The virtual one comes from
the two Born-like FFs with cluster index {\sf i=1}. It is proportional to the Born width
and contains the infrared divergence parameterized by the photon mass $\lambda$:
\bq
\frac{d^2\Gamma^{\rm Virt}}{ds\,d\cos\vartheta_f} =  
\frac{d^2\Gamma^{\rm Born}}{ds\,d\cos\vartheta_f}\frac{\alpha}{\pi}\qel^2  
\Biggl\{-\ln\left(\frac{\mf^2}{\lambda^2}\right)
\left[\ln\left(\frac{s}{\mf^2}\right)-1\right]
+\frac{1}{2}\ln\left(\frac{s}{\mf^2}\right)\left[3-\ln\left(\frac{s}{\mf^2}\right)\right]
+4\Litwo(1)-2
\Biggl\}.
\eq
The soft photon contribution is also proportional to the Born one;
 its infrared divergence 
cancels against  the virtual contribution. It contains also a logarithm with
 {\it soft-hard separator}
$\bar{\omega}$:
\bq 
\frac{d^2\Gamma^{\rm Soft}}{ds\,d\cos\vartheta_f} =   
\frac{d^2\Gamma^{\rm Born}}{ds\,d\cos\vartheta_f}
\frac{\alpha}{\pi} \qel^2
\Biggl\{\left[\ln\left(\frac{\mf^2}{\lambda^2}\right)
        +2\ln\left(\frac{2\bar{\omega}}{\mf}\right)
        -\ln\left(\frac{s}{\mf^2}\right)\right]
\left[\ln\left(\frac{s}{\mf^2}\right)-1\right]
-\Litwo\left(1\right)+1\Biggr\}.
\eq
The hard photon contribution after integration over three kinematical variables
$d\varphi_4\,,\;d\cos\vartheta_4$ and $d\tau$ (the first two vary together in full angular
$4\pi$ limits and $m^2_{f}\leq\tau\leq(\sqrt{s}-m_{f})^2$) is 
\bqa 
\nll\frac{d^2\Gamma^{\rm Hard}}{ds\,d\cos\vartheta_f} &=&
\frac{d^2\Gamma^{\rm Born}}{ds\,d\cos\vartheta_f}\frac{\alpha}{\pi}\qel^2
\Biggl\{-2\ln\left(\frac{2\bar{\omega}}{\mf}\right)\hspace*{-1mm}
\left[\ln\left(\frac{s}{\mf^2}\right)-1\right]
    -\frac{1}{2} \ln\left(\frac{s}{\mf^2}\right)\hspace*{-1mm}
\left[5-3 \ln\left(\frac{s}{\mf^2}\right)\right]
\nll &&\hspace*{9mm}
-3 \Litwo(1)+\frac{1}{4}\Biggr\}
+ k_{\sss B}\frac{\alpha}{\pi}\qel^2
\left[ \left(1-\frac{s}{\mh^2} \right)^2+10 s \frac{\mz^2}{\mh^4}-1
   +\left(1-\frac{\mz^2}{\mh^2}\right)^2 \right].
\eqa
The total QED correction, sum of above three contributions, is free not only of 
infrared divergence and of soft-hard separator, but also free of final fermion mass
singularity in accordance with the KLN
 theorem~\cite{Kinoshita:1962ur}--\cite{Lee:1964is}.
\bq
\frac{d^2\Gamma^{\rm Total}}{ds\,d\cos\vartheta_f} =
\frac{d^2\Gamma^{\rm Born}}{ds\,d\cos\vartheta_f}\frac{\alpha}{\pi}\qel^2
+ k_{\sss B} \frac{\alpha}{\pi}\qel^2\left[ \left( 1-\frac{s}{\mh^2} \right)^2 
+ 10 s \frac{\mz^2}{\mh^4} - 1 +\left(1-\frac{\mz^2}{\mh^2}\right)^2 \right].
\eq
Finally, if one integrates over $d\cos\vartheta_f$, the well known  $Z$ decay correction
factor restores:
\bq
\frac{d\Gamma^{\rm Total}}{ds} =
\frac{d\Gamma^{\rm Born}}{ds}\left[1+\frac{3}{4}\frac{\alpha}{\pi}\qel^2\right].
\eq
Therefore, the QED part of the correction is small, $\sim 0.2\%$.


\section{Numerical results and comparison\label{numerics}}
In the numerical calculations by {\tt s2n} package we use two precompiled libraries:
{\tt SancLib\_v1.00} and {\tt looptools 2.1}~\cite{Hahn:1998yk}.
\subsection{Numerical results for Electroweak corrections\label{ewc}}

\subsubsection{Process $f_1 \bar{f}_1\to HZ$}
For this process we present in Table 1 the results of a tuned triple comparison of
the one-loop electroweak corrections, excluding the gauge-invariant QED subset of diagrams
(vertex and electron self-energy) and real bremsstrahlung.
The input parameters are taken as in~\cite{Denner:1992bc}. Table 1 shows 6-7 digits 
agreement between the three calculations.
\begin{table}[!h]
\begin{center}
\begin{tabular}{||c|c|r|r|r||}
 \hline
$\sqrt{s}$, GeV&$M_H, GeV$
&\cite{Denner:1992bc}~~~~~&{\itshape Grace-Loop}
                                     & {\tt SANC}~~~~\\ 
\hline
$ 500$ & $100$ &    4.1524 &    4.15239 &    4.15239 \\
$ 500$ & $300$ &    6.9017 &    6.90166 &    6.90166 \\
$1000$ & $100$ &$-$ 2.1656 &$-$ 2.16561 &$-$ 2.16560 \\
$1000$ & $300$ &$-$ 2.4995 &$-$ 2.49949 &$-$ 2.49949 \\
$1000$ & $800$ &   26.1094 &   26.10942 &   26.10942 \\
$2000$ & $100$ &$-$11.5414 &$-$11.54131 &$-$11.54136 \\
$2000$ & $300$ &$-$12.8226 &$-$12.82256 &$-$12.82256 \\
$2000$ & $800$ &   11.2468 &   11.24680 &   11.24680 \\
\hline
\end{tabular}
\caption{\label{eezhcompar}
{Comparison of percentage correction to the total cross section
$e^+ e^-\to Z H$ between $\;$Ref.~\cite{Denner:1992bc}, {\itshape Grace-Loop}~\cite{Belanger:2003sd}
and {\tt SANC}.}}
\end{center}
\vspace*{-5mm}
\end{table}

\noindent
Beside the input given in \cite{Denner:1992bc}, $M_W$ is crucial for a precise comparison.
The following $M_W$ masses have been used: $M_W=80.231815$GeV ($M_H=100$GeV), $M_W=80.159313$
GeV ($M_H=300$GeV), $M_W=80.081409$GeV ($M_H=800$GeV), following
A.~Denner private communication, as referred to in~\cite{Belanger:2003sd}.

\subsubsection{Process $f_1 \bar{f}_1\to ZZ$}
For this process we compared only ``virtual~+~soft'' corrections in the conditions of Tables~1,2
of~Ref.\cite{Denner:1988tv} with input parameters tuned carefully. In this
case we do not find a good agreement: {\tt SANC} numbers happened to lie about 10\% lower.
We shall go back to searching for the origin of this discrepancy after a tuned 
comparison of the very similar process $f_1 \bar{f}_1\to ZA$ with the results 
of~Ref.\cite{Bohm:1986mz}.
The implementation of the latter process into the {\tt SANC} system is nearly finished.

\subsection{Numerical results for real and complete corrections\label{real}}

\subsubsection{Hard bremsstrahlung in $f_1\bar{f}_1\to HZ$ annihilation channel}
In  Table~\ref{comparisoneeHZA} we present typical results of a triple comparison of the 
Born cross section and the cross section of hard photon bremsstrahlung between two calculations 
within {\tt SANC} (semi-analytic, Eq.\ref{Hardann}, and MC) and those of CompHEP for 
$E_\gamma\geq 1$GeV. Here we used $m_H$=130 GeV and the other parameters as in CompHEP.

\begin{table}[!h]
\begin{center}
\begin{tabular} {||l|l|l|l|l|l||}
\hline
\hline
&\multicolumn{5}{|c||}{$\sigma$, pb}                                                      \\
\hline
$\sqrt{s}$, GeV   &$    250    $&$    300    $&$   500     $&$   1000    $&$    2000    $ \\
\hline
Born (SANC)       &$0.21984(1) $&$0.17454(1) $&$0.056890(1)$&$0.012898(1)$&$0.0031322(1)$ \\
Born (CompHEP)    &$0.21984(1) $&$0.17454(1) $&$0.056889(1)$&$0.012898(1)$&$0.0031322(1)$ \\
\hline
Hard (SANC,~s2n)  &$0.080309(1)$&$0.091168(1)$&$0.043650(1)$&$0.013246(1)$&$0.0040293(1)$ \\
Hard (SANC,~MC)   &$0.080307(1)$&$0.091166(1)$&$0.043649(1)$&$0.013246(1)$&$0.0040293(1)$ \\
Hard (CompHEP)    &$0.080306(2)$&$0.091168(2)$&$0.043651(1)$&$0.013242(3)$&$0.0040287(4)$ \\
\hline 
\hline
\end{tabular}
\end{center}
\vspace*{-2mm}
\caption{Comparison of the Born cross section and hard photon cross section
of $e^+e^- \to HZ\gamma$ reaction for $E_\gamma\geq 1$\,GeV.}
\label{comparisoneeHZA}
\vspace*{-2mm}
\end{table}

The two {\tt SANC} results perfectly agree within statistical errors. The agreement with 
CompHEP also looks quite good.

\subsubsection{Hard bremsstrahlung in $f_1\bar{f}_1\to ZZ$ annihilation channel}
In  Table~\ref{comparisoneeZZA} we present the results of a double comparison of the 
Born cross section and the cross section of hard photon bremsstrahlung between the calculation
within {\tt SANC} (MC) and those of CompHEP for $E_\gamma\geq 1$GeV. 
\footnote{For this channel we don't have the semi-analytic result.}
Here we used $m_H$=130 GeV and the other parameters as in CompHEP.

\begin{table}[!h]
\begin{center}
\begin{tabular} {||l|l|l|l|l|l||}
\hline
\hline
\multicolumn{6}{|c||}{  $\sigma$, pb}
\\
\hline
$\sqrt{s}$,GeV &$   250     $&$   300   $&$   500     $&$   1000     $&$ 2000     $ \\
\hline
Born (SANC)    &$1.0758(1) $&$0.82971(1)$&$0.40644(1)$&$0.14815(1)$&$0.049760(1)$ \\
Born (CompHEP) &$1.0758(1) $&$0.82971(1)$&$0.40644(1)$&$0.14814(1)$&$0.049761(1)$\\
\hline
Hard (SANC,~MC)&$0.52262(1)$&$0.46800(2)$&$0.29114(1)$&$0.13375(1)$&$0.54563(1)$ \\ 
Hard (CompHEP) &$0.52258(2)$&$0.46801(1)$&$0.29111(2)$&$0.13374(1)$&$0.54563(6)$ \\
\hline
\hline
\end{tabular}
\end{center}
\vspace*{-2mm}
\caption{Comparison of the Born cross section and hard photon cross section
$e^+e^- \to ZZ\gamma$ reaction for $E_\gamma\geq 1$\,GeV.}
\label{comparisoneeZZA}
\vspace*{-2mm}
\end{table}

Again, we see a very good within larger statistical errors in CompHEP. 

\clearpage

\subsubsection{$H\to f_1\bar{f}_1 Z$ decay channel}
For this decay we present the complete one-loop correction.

We present numbers, collected in the $G_\mu$ scheme for the standard {\tt SANC} INPUT: 
PDG(2006)~\cite{PDG2006}
\bqa \nonumber
\begin{array}[b]{lcllcllcllcllcl}
G_{\sss F} & = & 1.16637\cdot 10^{-5}\GeV^{-2}, &\alpha(0) &=& 1/137.03599911, 
&\alpha_s(\mz) &=& 0.1187,\\
\mw & = & 80.403\;\GeV, &
\gw & = &  2.141\;\GeV, \\
\mz & = & 91.1876\;\GeV,& 
\gz & = & 2.4952\;\GeV, \\
\mh & = & 120\;\GeV,    &\\
m_e & = & 0.51099892\cdot 10^{-3}\;\GeV,& m_u & = & 62\;\MeV & m_d & = & 83\;\MeV,\\
m_{\mu}&=&0.105658369\;\GeV,& m_c & = &1.5\;\GeV, & m_s & = & 215\;\MeV,          \\
m_{\tau}&=&1.77699\;\GeV,   & m_b & = &4.7\;\GeV, & m_t & = & 174.2\;\GeV,        \\
\Gamma_t&=&1.551\;\GeV.\\
\end{array}
\eqa
The only exception is the Higgs boson mass for which we again take $\mh=130\;\GeV$ 
in this section.

Table~\ref{rezHffZ2} shows the double and singly differential decay width for the decays 
$H\to e^{+}e^{-}Z$ and $H\to \mu^{+}\mu^{-} Z$ for a set of $s$ and $\cos{\vartheta_{l}}$.

From Table~\ref{rezHffZ2} it is  seen that at the edges of $\cos{\vartheta_{l}}$ and 
near the fermionic threshold the double differential width shows a $1/s$ behaviour, typical 
for Coulomb interaction. 
The origin of the Coulomb peak at the one-loop level may be easily understood.
First we note that $H\to Z\gamma$ width does not vanish for an on-shell photon with 
$Q^2_{\gamma}=0$, see first Fig.~\ref{compton}:


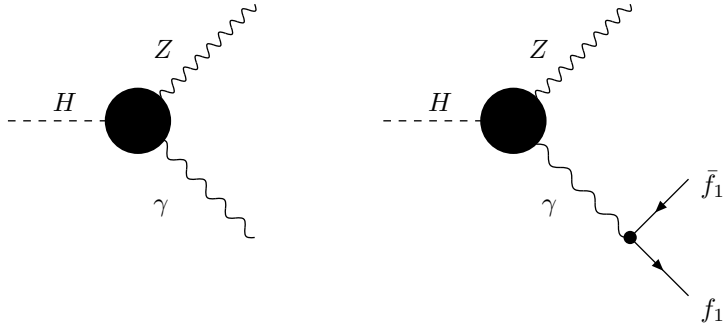
\begin{figure}[!h]
\vspace*{-10mm}
\[
\begin{array}{cc}
\begin{picture}(132,132)(0,0)
 \DashLine(-5,66)(44,66){3}
 \Vertex(44,66){12.5}
 \Text(12, 70)[lb]{$H$}
\Photon(44,66)(88,110){2}{10}
\Photon(44,66)(88,22){2}{6}
 \Text(50,90)[lb]{$Z$}
 \Text(50,30)[lb]{$\gamma$}
\end{picture}
&
\begin{picture}(132,132)(0,0)
 \DashLine(-5,66)(44,66){3}
 \Vertex(44,66){12.5}
 \Text(12, 70)[lb]{$H$}
\Photon(44,66)(88,110){2}{10}
\Photon(44,66)(88,22){2}{5}
 \Vertex(88,22){2.5}
 \Text(50,90)[lb]{$Z$}
\ArrowLine(110,44)(88,22)
\ArrowLine(88,22)(110,0)
 \Text(115,37)[lb]{$\bar{f}_1$}
 \Text(115,-10)[lb]{$f_{1}$}
 \Text(55,30)[lb]{$\gamma$}
\end{picture}
\end{array}
\]
\vspace*{-5mm}
\caption[$H\to Z\gamma$ decay and Compton singularity\label{compton}]
        {$H\to Z\gamma$ decay and Compton singularity\label{compton}.} 
\end{figure}

\noindent
Therefore, the one-loop amplitude for $H\to Z f_1\bar{f}_1$ with virtual photon exchange 
will show a  $\sim 1/s$ behaviour (with $s=-Q^2_{\gamma}$). This, in turn, will lead 
to the $\sim 1/s$ behaviour of both the double and single decay differential widths. 
This conclusion is fully confirmed by the numbers in Table~\ref{rezHffZ2}.

Recalling now the limits of $s, \; 4 m^2_{f} \leq s \leq (\mh-\mz)^2$, one might expect
the appearance after integration over $s$ of the big logarithm $\ln((\mh-\mz)^2/m^2_{f})$,
with a  final state fermion mass singularity. However, the $1/s$ region is
 very narrow and it is
largely washed out not only by a soft cut on the variable $s$ but even by the plain
integration over $s$. 

\vspace*{5mm}
\begin{table}[h!]
\begin{center}
\begin{tabular} {||c|c||c|c|c|c||}
\hline\hline
\multicolumn{6}{||c||}
{$H\to e^{+}e^{-} Z$}\\
\hline\hline
\multicolumn{6}{||c||}{\phph$\qquad$ Part 1, $d^2\Gamma/ds\,d\cos{\vartheta_{l}}\cdot10^8$, GeV$^{-1}$}\\
\hline
&$\sqrt{s}$, GeV                          &$     1   $&$    3    $&$     10  $&$   38    $ \\
\hline
Born      &                               &$ 0.04760 $&$ 0.04910 $&$ 0.06596 $&$ 0.09604 $ \\
1-loop    & $\cos{\vartheta_{l}}=\pm 0.9$ &$ 0.54806 $&$ 0.10413 $&$ 0.07078 $&$ 0.09875 $ \\
$\delta$  &                               &$ 10.5123 $&$ 1.12053 $&$ 0.07301 $&$ 0.02829 $ \\
\hline      
Born      &                               &$ 0.18725 $&$ 0.18788 $&$ 0.19473 $&$ 0.09768 $ \\
1-loop    & $\cos{\vartheta_{l}}=\pm 0.5$ &$ 0.53457 $&$ 0.22763 $&$ 0.20006 $&$ 0.10044 $ \\
$\delta$  &                               &$ 1.85478 $&$ 0.21159 $&$ 0.02737 $&$ 0.02823 $ \\
\hline
Born      &                               &$ 0.24960 $&$ 0.24983 $&$ 0.25221 $&$ 0.09842 $ \\
1-loop    & $\cos{\vartheta_{l}}=\ph 0.0$ &$ 0.52882 $&$ 0.28303 $&$ 0.25801 $&$ 0.10120 $ \\
$\delta$  &                               &$ 1.11867 $&$ 0.13289 $&$ 0.02296 $&$ 0.02820 $ \\
\hline
\hline
\multicolumn{6}{||c||}{ \phph Part 2, $d\Gamma/ds \cdot 10^{9}$, GeV$^{-1}$} \\
\hline
\multicolumn{2}{||c||}{Born}              &$ 3.330 $&$ 3.345 $&$ 3.511 $&$ 1.949 $\\
\multicolumn{2}{||c||}{1-loop}            &$ 10.73 $&$ 4.186 $&$ 3.618 $&$ 2.004 $\\
\multicolumn{2}{||c||}{1-loop/Born}       &$ 2.224 $&$ 0.252 $&$ 0.030 $&$ 0.028 $\\
%
\hline\hline
\multicolumn{6}{||c||}
{$H\to \mu^{+}\mu^{-} Z$}\\
\hline\hline
\multicolumn{6}{||c||}{\phph$\qquad$ Part 1, $d^2\Gamma/ds\,d\cos{\vartheta_{l}}\cdot10^8$, GeV$^{-1}$}\\
\hline
Born      &                               &$ 0.05533 $&$ 0.04996 $&$ 0.06602 $&$ 0.09603 $ \\
1-loop    & $\cos{\vartheta_{l}}= - 0.9$  &$ 0.54437 $&$ 0.10480 $&$ 0.07081 $&$ 0.09875 $ \\
$\delta$  &                               &$ 8.83770 $&$ 1.09736 $&$ 0.07265 $&$ 0.02829 $ \\
\hline
Born      &                               &$ 0.18573 $&$ 0.18770 $&$ 0.19470 $&$ 0.09768 $ \\
1-loop    & $\cos{\vartheta_{l}}= - 0.5$  &$ 0.52536 $&$ 0.22739 $&$ 0.20003 $&$ 0.10044 $ \\
$\delta$  &                               &$ 1.82850 $&$ 0.21142 $&$ 0.02740 $&$ 0.02823 $ \\
\hline
Born      &                               &$ 0.24395 $&$ 0.24919 $&$ 0.25214 $&$ 0.09842 $ \\
1-loop    & $\cos{\vartheta_{l}}=\ph 0.0$ &$ 0.51713 $&$ 0.28238 $&$ 0.25795 $&$ 0.10044 $ \\
$\delta$  &                               &$ 1.11982 $&$ 0.13317 $&$ 0.02303 $&$ 0.02823 $ \\
\hline
Born      &                               &$ 0.18573 $&$ 0.18770 $&$ 0.19470 $&$ 0.09768 $ \\
1-loop    & $\cos{\vartheta_{l}}=\ph 0.5$ &$ 0.52538 $&$ 0.22739 $&$ 0.20003 $&$ 0.10044 $ \\
$\delta$  &                               &$ 1.82862 $&$ 0.21144 $&$ 0.02740 $&$ 0.02823 $ \\
\hline
Born      &                               &$ 0.05533 $&$ 0.04996 $&$ 0.06602 $&$ 0.09603 $ \\
1-loop    & $\cos{\vartheta_{l}}=\ph 0.9$ &$ 0.54441 $&$ 0.10480 $&$ 0.07081 $&$ 0.09875 $ \\
$\delta$  &                               &$ 8.84841 $&$ 1.09749 $&$ 0.07266 $&$ 0.02829 $ \\
\hline
\hline
\multicolumn{6}{||c||}{\phph Part 2, $d\Gamma/ds \cdot 10^{9}$, GeV$^{-1}$} \\
\hline
\multicolumn{2}{||c||}{Born}              &$ 3.327 $&$ 3.344 $&$ 3.511 $&$ 1.949 $\\
\multicolumn{2}{||c||}{1-loop}            &$ 10.57 $&$ 4.184 $&$ 3.617 $&$ 2.004 $\\
\multicolumn{2}{||c||}{1-loop/Born}       &$ 2.176 $&$ 0.251 $&$ 0.030 $&$ 0.028 $\\
\hline
\hline
\end{tabular}
\end{center}
\vspace*{-3mm}
\caption{The double and single differential widths for two decay channels: $H\to e^{+}e^{-} Z$ and
$H\to \mu^{+}\mu^{-} Z$. The table contains two parts for each channel.
Part 1: first row: the double differential decay width 
$d^2\Gamma/ds\,d\cos{\vartheta_{l}}\cdot 10^{8} $GeV$^{-1}$ at the Born level;
second row: the  double differential decay width at the 1-loop level;
third row: relative correction $\delta=d^2\Gamma^{\rm{1-loop}}/d^2\Gamma^{\rm{Born}}$.
Numerical values are truncated to 6 figures.
Part 2: the same set for the single decay differential width 
$d\Gamma/ds \cdot 10^{9}$, GeV$^{-1}$.
Numerical values are rounded.
\label{rezHffZ2}}
\vspace*{-4mm}
\end{table}
Finally let us discuss the total width for muon channel. In Born approximation 
it is
$\Gamma^{\rm Born}=5.592\cdot 10^{-6}$ GeV,
while with the complete EW
corrections it is
$\Gamma^{\rm Born+1-loop}=5.774\cdot 10^{-6}$ GeV.
So, the correction in $G_\mu$ scheme amounts to $3.2\%$.
\vspace*{-5mm}

\clearpage 

\subsubsection{Hard photon radiation in $H\to f_1\bar{f}_1 Z$ decay}
Here we present the triple comparison of the hard photon bremsstrahlung contribution with a cut on
the lepton--photon invariant mass $\tau=-(p_4+p_5)^2$ again between two calculations of {\tt SANC} 
and one of CompHEP.

\begin{table}[!h]
\begin{tabular}{|c|l|l|l|l|l|l|}
\hline
\multicolumn{7}{|c|}{ $\Gamma\cdot 10^{7}$,~GeV}                                           \\
\hline
$\sqrt{\tau_{min}}$ &  $\sqrt{\tau_{min}}_{|\bar\omega =0.1}$GeV
                             &$ 1       $&$   2      $&$   3     $&$    5    $&$   10 $    \\[2mm]
\hline
\multicolumn{7}{|c|}{$e$}                                                                  \\
\hline
s2n     &$57.456 $           &$17.982   $&$ 13.094   $&$10.395   $&$7.2097   $&$3.4621    $\\
MC      &$57.5(2)$           &$17.981(3)$&$ 13.098(2)$&$10.398(1)$&$7.2100(5)$&$3.4619(7) $\\
CompHEP &$55.3(5)$, unstable &$17.96(1) $&$ 13.10(1) $&$10.39(1) $&$7.198(5) $&$3.458(2)  $\\
\hline
\multicolumn{7}{|c|}{$\mu$}                                                                \\
\hline
s2n     &$18.917   $         &$9.6862   $&$ 6.8385   $&$5.3224   $&$3.5885   $&$1.6396    $\\
MC      &$18.916(3)$         &$9.6869(6)$&$ 6.8384(5)$&$5.3227(4)$&$3.5886(8)$&$1.6399(3) $\\
CompHEP &$18.91(1) $         &$9.678(7) $&$ 6.837(5) $&$5.323(4) $&$3.586(3) $&$1.646(1)  $\\
\hline
\hline
\multicolumn{7}{|c|}{$\tau$}                                                               \\
\hline
   s2n       &$5.9714   $    &  out of   &$ 5.0085   $&$2.9539   $&$1.7377   $&$0.68272   $\\
   MC        &$5.9717(4)$    &kinematical&$ 5.0081(5)$&$2.9542(3)$&$1.7375(1)$&$0.68269(5)$\\
   CompHEP   &$5.983(4) $    & region    &$ 5.016(4) $&$2.959(2) $&$1.741(1) $&$0.6832(5) $\\
\hline
\hline
\end{tabular}
\caption[The decay width, for three channels: $e,\mu,\tau$]
{The decay width $\Gamma$ in GeV, massive case, as
 a function of $\sqrt{\tau_{min}}$ from {\tt SANC} s2n, {\tt SANC} MC and CompHEP\label{hardtau}}
\end{table}
In  Table~\ref{hardtau} ``$\sqrt{\tau_{min}}_{|\bar\omega =0.1}$GeV'' denotes $\tau_{min}$ derived 
by the formula $\tau_{min}=m^2_l+2m_l E_{\gamma,min}$.\\
As seen, two {\tt SANC} numbers agree within MC errors and there is a reasonable agreement with 
CompHEP everywhere but upper left corner (soft radiation by electrons) where CompHEP show a tendency
to be unstable. 


\section{A Monte - Carlo generator for  $H \to 4 \mu$\label{generator}}
A Monte Carlo generator of unweighted events for process $H \to 4 \mu$ is the first 
example of a possible application of the building blocks ideology of {\tt SANC}. 

\begin{floatingfigure}{80mm}
\includegraphics[width=6cm,height=9cm]{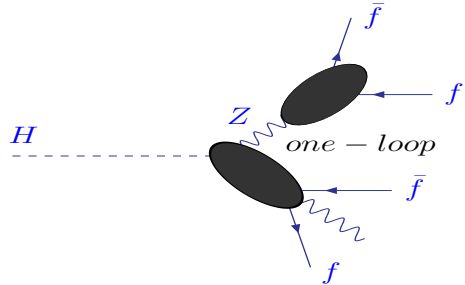}
\vspace*{-52mm}
\caption[$H \to 4 \mu$ in the single resonance \\ approximation.]
{$H \to 4 \mu$ in the single resonance \\ approximation.\label{sresa}}
\vspace*{-1mm}
\end{floatingfigure}

\noindent
In our generator we implement two building blocks at the one-loop level:
$H \to ff Z(\gamma)$ and $Z \to ff(\gamma)$.
We merge these two blocks and create a link between them by means of the
$Z^{*}$ (resonating boson) line with Breit-Wigner mass distribution. 
In this spirit we create the generator in the single resonance approximation.
We also built a double resonance generator, where we incorporated resonance approximation 
in two $Z^{*}$ lines.

 The range for application of the single resonance approximation was found to be
$120$ GeV$\leq\mh\leq 160$ GeV and for the double resonance approximation $\mh\geq$ 180 GeV.

\begin{figure}[!ht]
\includegraphics[width=15cm,height=8cm]{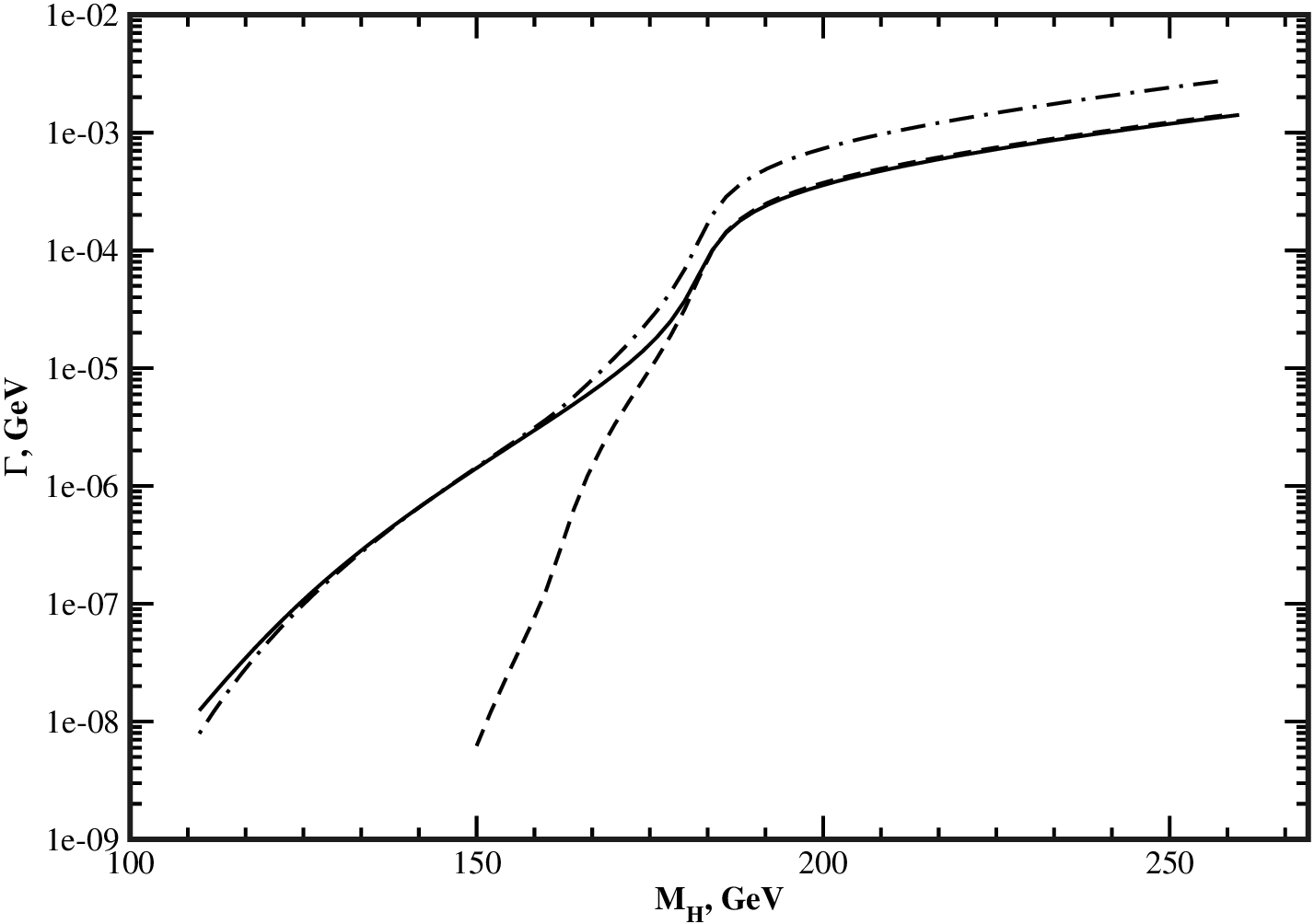}
\vspace*{-3mm}
\caption[Born width of the decay $H \to 4 \mu$ in three approximations]
{Born width of the decay $H \to 4 \mu$ in three approximations.\label{justsresa}}
\end{figure}

These conclusions are illustrated by Fig.~\ref{justsresa} where we present the results of 
calculations of the tree level width of the decay $H \to 4\mu$ for the three cases:\\[1mm]
1) solid line: results of complete tree level calculations neglecting effects of identical 
 final state muons; \\
2) dash dotted line: single resonance approximation,
\bqa
\Gamma^{\rm {1-res}}_{H\to 4\mu}=\frac{\Gamma_{H\to 2\mu Z}\Gamma_{Z\to 2\mu}}{\Gamma_Z}\;;
\eqa
3) dashed line: double resonance approximation,
\bqa
\Gamma^{\rm {2-res}}_{H\to 4\mu}=\frac{\Gamma_{H\to 2 Z}\Gamma^2_{Z\to 2\mu}}{\Gamma^2_Z}\;.
\eqa
\vspace*{6mm}

The loop-corrected result is the linear combination of three types of events:
Born with Identity events minus Resonanse Born events plus One-loop Resonance events.
We feel that this is the main shortcoming of the generator. 

Let us consider each type of events:
\begin{itemize}
\vspace*{-1mm}
\item{\underline{``Born with Identity'' events} means a branch which computes the distributions 
without radiative corrections but with effects of identical muons;}
\vspace*{-1mm}
\item{\underline{``Resonance Born'' events} means resonance approximation for one of the $Z$ 
bosons, i.e. $H \to Z Z^{*} \to \mu^+\mu^-\mu^+\mu^-$, where $Z^*$ is the resonating $Z$. 
Here, the two building blocks in Fig.~\ref{sresa} are calculated at the Born level;}
\vspace*{-1mm}
\item{\underline{``Resonance One-loop'' events} is implemented in the same spirit as 
``Resonance Born'', but building blocks $H \to ff Z(\gamma)$ and $Z \to ff(\gamma)$ are 
calculated at the one-loop level.}
\end{itemize}
The codes of both MC generators can be obtained from the authors by request.

Recently appeared a new MC code Prophecy4f, see 
Refs.\cite{Bredenstein:2006rh}--\cite{Buttar:2006zd}, realizing calculation of the complete 
one-loop corrected partial widths of the $H\to 4l$ channels. 
We present a preliminary comparison between MC Prophecy4f and {\tt SANC} in
Table~\ref{ComparisonwithDenner}.

\begin{table}[h!]
\begin{center}
\begin{tabular}{||c|c|c|c|c|c||}
\hline
$\sqrt{s}$, GeV & $120$ & $130$ & $140$ & $150$ & $160$
\\ \hline  
  Prophecy4f    & $7.053(3) \cdot 10^{-8}$
                & $2.3769(9)\cdot 10^{-7}$
                & $6.692(2) \cdot 10^{-7}$
                & $1.6807(6)\cdot 10^{-6}$
                & $4.006(1) \cdot 10^{-6}$\\ \hline
SANC ($G_{\mu}$)& $7.197(3) \cdot 10^{-8}$ 
                & $2.4079(8)\cdot 10^{-7}$ 
                & $6.743(2) \cdot 10^{-7}$ 
                & $1.6842(5)\cdot 10^{-6}$ 
                & $3.962(2) \cdot 10^{-6}$ 
                                          \\ \hline
$\delta, \%$  & 2.04 & 1.01 & 0.76 & 0.21 & -1.10\\ \hline
SANC ($\alpha$) & $6.938(2) \cdot 10^{-8}$ 
                & $2.343(1) \cdot 10^{-7}$ 
                & $6.594(2) \cdot 10^{-7}$ 
                & $1.6534(5)\cdot 10^{-6}$ 
                & $3.915(1) \cdot 10^{-6}$ 
                                          \\ \hline
\end{tabular}
\caption{Comparison for partial width for decay $H \to 4 \mu$
in $ G_\mu$ scheme for $\mh=140$ GeV between Prophecy4f and {\tt SANC}.}
\label{ComparisonwithDenner}
\end{center}
\end{table}
\vspace*{-3mm}

As seen from the Table, there is $\pm$1\% agreement in the mass range 130--140 GeV,
degrading at the edges of the interval [120--160], that finds
its natural explanation in Fig.\ref{justsresa}. Moreover, Prophecy4f uses the
complex-mass scheme and takes into account several higher order corrections.
One has to emphasize, however, that {\tt SANC} calculations in $\alpha$ and $G_\mu$ schemes 
differ by about 2\%. This can be considered as a rough estimate of the theoretical error.
Prophecy4f numbers lie basically inside the range of SANC predictions.

The generator in the single resonance approximation described in this section was used
for a MC simulation of $H\to 4\mu$ decay in the ATLAS detector and the results were compared 
with simulation by PYTHIA, showing notable deviations, see~\cite{Boyko:2006April}.
This fact demonstrates the importance of higher order corrections and the necessity to reduce 
the theoretical error.

\section{User Guide}\label{Uguide}
\subsection{Benchmark case 3: the process $H\to f_1 \bar{f_1} Z$\label{Uguide:hffz_decay}}
Here we consider the $2f2b$ {\bf NC} process $H\to f_1\bar{f}_1 Z$.

One can open the relevant branch of the {\tt SANC} tree as follows:\\[3mm]
\centerline{\bf EW $\to$ Processes $\to$ 4 legs $\to$ 2f2b $\to$ Neutral Current $\to$ H$\to$f1f1Z}\\[3mm]
For this process there are three FORM programs: ({\bf{FF}}) {\it Form Factors}, 
({\bf{HA}}) {\it Helicity Amplitudes}, and ({\bf{BR}}) {\it Brem\-sstrah\-lung}.
Each of them in turn is opened, compiled and run as described in
Section 6 of Ref.\cite{Andonov:2004hi}.

For the process $H\to e^+ e^- Z$ we have in the {\bf Console} window the particle
indices shown in Table~\ref{hffz}.
\begin{table}[!h]
\begin{center}
\caption{Assignment of particle indices for the process  $H\to e^+ e^- Z$}\label{hffz}
\vspace*{3mm}
\begin{tabular}{||c|l||} 
\hline\hline
typeIU = 4\phantom{4}  & initial 
partile (H-boson) \\
typeID = 12            & final particle (electron)         \\
typeFU = 12            & final antiparticle (positron)     \\
typeFD = 2\phantom{2}  & final particle (Z-boson)          \\ 
\hline\hline
\end{tabular}
\vspace*{-3mm}
\end{center}
\end{table}

These can be changed to typeID (typeFU) = 13,14 for up- and down-quarks
in the final state of the processes $H\to (u {\bar u}, d {\bar d}) Z$
by editing the particle numbers as explained in Section 6 of Ref.\cite{Andonov:2004hi}.
\footnote{See Table 2 for definitions of particle types {\sf typeXX}.}.

Next bring the {\it Fortran Editor} sheet of the {\bf Editors List}
and the {\bf Numeric Form} panel to the foreground.
Shown in the {\bf Numeric Parameter} sheet are the particle masses
in GeV and the invariant mass of $f{\bar f}Z$ compaund in GeV,
also the cosine of the angle $\vartheta_l$ defined in Fig.\ref{Blevel}.

Click on the {\bf Rehash} button at the bottom of the 
{\bf Numeric Form} panel: the main module of FORTRAN code appears in the
{\it Fortran Editor} sheet of the {\bf Editors List}.
Then click on {\bf Compile}. The final answer appears in the {\bf Output} field.
It consists of the parameters used ($\alpha$, $G_F$, particle masses,
the 't Hooft scale $\mu$ and the invariant mass of compaund),
and the resulting differential width
$d^2\Gamma/ds\,d\cos\vartheta_l$ in the {\sf Born} approximation
and  {\sf Born+one-loop}. The results for the default parameters
and for several scattering angles are summarised in Table~\ref{rezHffZ3}.
\vspace*{-3mm}

\begin{table}[h!]
\caption{The double differential widths for $H\to e^{+}e^{-} Z$ channel in $\alpha$-scheme:
first row: the double differential decay width 
$d^2\Gamma/ds\,d\cos{\vartheta_{l}}\cdot 10^{8} $GeV$^{-1}$ at the Born level;
second row: the  double differential decay width at the 1-loop level;
third row: relative correction $\delta=d^2\Gamma^{\rm{1-loop}}/d^2\Gamma^{\rm{Born}}$.
Numerical values are truncated to 6 figures.
\label{rezHffZ3}}
\begin{center}
{\small
\begin{tabular} {||c|c||c|c|c|c||}
\hline\hline
\multicolumn{6}{||c||}
{$ H\to e^{+}e^{-} Z$}\\
\hline\hline
\multicolumn{6}{||c||}{\phph$\qquad$ Part 1, $d^2\Gamma/ds\,d\cos{\vartheta_{l}}\cdot10^8$, GeV$^{-1}$}\\
\hline
&$\sqrt{s}$, GeV                           &$      1   $&$    3    $&$     10  $&$    28   $\\
\hline
Born       &                               &$  0.02019 $&$ 0.02144 $&$ 0.03505 $&$ 0.04261 $ \\
1-loop     &$\cos{\vartheta_{l}}=\pm 0.9$ &$  0.21060 $&$ 0.04321 $&$ 0.03874 $&$ 0.04602 $ \\
$\delta $  &                               &$  9.43022 $&$ 1.01508 $&$ 0.10537 $&$ 0.07984 $ \\
\hline
Born       &                               &$  0.07914 $&$ 0.07964 $&$ 0.08478 $&$ 0.04353 $ \\
1-loop     & $\cos{\vartheta_{l}}=\pm 0.5$ &$  0.21495 $&$ 0.09898 $&$ 0.09150 $&$ 0.04701 $ \\
$\delta $  &                               &$  1.71589 $&$ 0.24281 $&$ 0.07922 $&$ 0.07976 $ \\
\hline      
Born       &                               &$  0.10546 $&$ 0.10562 $&$ 0.10698 $&$ 0.04394 $ \\
1-loop     & $\cos{\vartheta_{l}}=\ph 0.0$ &$  0.21695 $&$ 0.12394 $&$ 0.11510 $&$ 0.04745 $ \\
$\delta $  &                               &$  1.05716 $&$ 0.17343 $&$ 0.07586 $&$ 0.07972 $ \\
\hline
\hline
\end{tabular}
}
\end{center}
\end{table}

Input parameters can be changed by editing the appropriate field of
the  {\bf Numeric Form} panel and pressing the {\bf Rehash} button.
Again the {\bf Rehash} button must be pressed before pressing {\bf Compile}.

To produce whole Table~\ref{rezHffZ3} one can set flag
{\tt tbprint = 1} in the {\it Fortran Editors} sheet.
After editing the code just press {\bf Compile}; there is no need to press
the {\bf Rehash} button.

One can also produce the differential width $d\Gamma/ds$ and total width
$\Gamma$ in GeV by integrating the above differential width.
To produce these numbers one can set flag
{\tt inflag = 1,2}, respectevely, in the {\it Fortran Editors} sheet.
After editing the code just press {\bf Compile}; again, there is no need to press
the {\bf Rehash} button.

\vspace*{5mm}

\leftline{\Large\bf Acknowledgments}

\vspace*{2mm}

The authors are grateful to P.~Christova for a valuable discussion of bremsstrahlung issues,
to A.~Arbuzov for discussion of the stability of numerical calculations and physical results
and also to W.~Hollik for providing us with useful references.
Three of us (D.B., L.K. and G.N.) are cordially indebted to S.~Jadach and Z.~Was for offering
us an opportunity of encouraging common work at IFJ Krakow in April--May 2005 and to the IFJ
directorate for hospitality which was extended to us in this period,  
when the major part of this study was done.
We are thankful to the authors of the Prophecy4f generator for providing us with their numbers
for Table~\ref{ComparisonwithDenner}.

\clearpage

\def\href#1#2{#2}
\end{document}